\documentclass{iopart}
\usepackage{bbm,amssymb,amsbsy,times,pifont}
\usepackage[dvips]{graphicx}
\usepackage{hyperref}
\usepackage[latin1]{inputenc}
\usepackage{subfigure}
\usepackage{fancyhdr,makeidx}
\usepackage{color}

\newcommand{\prl}{{\it Phys. Rev. Lett.} }

\newcommand{\N}{\cal N}

\newcommand{\id}{\mathbbm{1}}

\renewcommand{\tr}{{\rm Tr}\,}
\renewcommand{\det}{{\rm Det}\,}
\newcommand{\gr}[1]{\boldsymbol{#1}}
\newcommand{\be}{\begin{equation}}
\newcommand{\ee}{\end{equation}}
\newcommand{\bea}{\begin{eqnarray}}
\newcommand{\eea}{\end{eqnarray}}
\newcommand{\ket}[1]{|#1\rangle}
\newcommand{\bra}[1]{\langle#1|}

\newcommand{\sig}{\gr{\sigma}}

\newcommand{\eq}[1]{Eq.~(\ref{#1})}

\begin{document}
\title[Manipulating the quantum information of the radial modes
of trapped ions]{Manipulating the quantum information of the
radial modes of trapped ions: Linear phononics, entanglement
generation, quantum state transmission and non-locality tests}
\author{Alessio Serafini,${}^{1,2}$ Alex Retzker,${}^{2,3}$ and Martin B.~Plenio${}^{2,3}$}
\address{${}^1$ Department of Physics \& Astronomy, University College London, Gower Street, London WC1E 6BT, UK\\
${}^2$ Institute for Mathematical Sciences, 53 Prince's Gate,
Imperial College London, London SW7 2PG, UK\\
${}^3$ QOLS, Blackett Laboratory, Imperial College London, London SW7 2BW, UK}
\date{24-09-2008}

\begin{abstract}
We present a detailed study on the possibility
of manipulating quantum information encoded in the ``radial''
modes of arrays of trapped ions ({\em i.e.}, in the ions'
oscillations orthogonal to the trap's main axis).
In such systems, because
of the tightness of transverse confinement, the radial modes
pertaining to different ions can be addressed individually.
In the first part of the paper we show that,
if local control of the radial trapping frequencies is
available, {\em any} linear optical and squeezing operation
on the locally defined modes -- on single as well as on many
modes -- can be reproduced by manipulating the frequencies.
Then, we proceed to describe schemes apt to generate
unprecedented degrees of bipartite and multipartite continuous
variable entanglement under realistic noisy working conditions, and
even restricting only to a global control of the trapping
frequencies. Furthermore, we consider the transmission of
the quantum information encoded in the radial modes along the
array of ions, and show it to be possible to a remarkable
degree of accuracy, for both finite-dimensional and continuous
variable quantum states. Finally, as an application, we
show that the states which can be generated in this setting
allow for the violation of multipartite non-locality tests,
by feasible displaced parity measurements. Such a demonstration
would be a first test of quantum non-locality for ``massive''
degrees of freedom ({\em i.e.}, for degrees of freedom
describing the motion of massive particles).
\end{abstract}
\maketitle

\section{Prologue: the promise of alternative continuous variable degrees of freedom}

The last decade saw a boom in the development of experimental
capabilities available for quantum information processing. The
ability to manipulate the information of discrete variables
encoded in polarisation, spin and internal atomic degrees of
freedom has by now reached very high standards. On the other
hand, the control and manipulation of continuous variable (CV)
quantum information is still almost exclusive to light fields
in quantum optical settings. Even though purely optical systems
rely on well established tools and are the natural choice for
communication tasks over long distances, they also suffer from
significant drawbacks.
Notably, the entanglement generation in such systems is strongly
limited by the efficiency of parametric processes in nonlinear
crystals; moreover, `static' optical degrees of freedom -- {\em
i.e.}~light resonating in cavities -- are seriously affected by
losses and decoherence over their typical dynamical time scales.

Such limitations motivate the question of whether the full potential
of infinite dimensional Hilbert spaces could be better harnessed by
``massive'' ({\em i.e.}, related to the position of a massive particle)
CV degrees of freedom.
Of course, to compete with the so far very successful
quantum optical toolbox, such degrees of freedom would have to allow
for a range of coherent manipulations at least as exhaustive
as what quantum optics currently permits.
Besides, to be considered advantageous over quantum optical
systems, such degrees of freedom should allow for
notable improvements in the generation of
quantum entanglement and squeezed states under realistic working conditions.

In the present paper, we argue that trapped ions meet both such requirements,
and present an extensive study to substantiate this argument.
In particular, we shall focus on the radial motion of trapped ions,
that is on oscillations along a direction orthogonal with respect to the
array of ions.
These oscillations are described by
continuous variable quantum degrees of freedom
which we will refer to as {\em radial modes} \cite{zho2006,nonlinear}.
Radial modes have attracted considerable interest in the last few years,
mostly in view of the fact that they allow for a tighter confinement
(which also permits one to define the phonons locally).

The paper is organised as follows.
After having introduced the description of the physical system (Sec.~\ref{trap}),
in Sec.~\ref{manip} we demonstrate that
any linear optical and squeezing operation can be obtained for radial modes
of trapped ions by controlling the individual {\em radial} trapping frequencies,
indicating that trapped ions can at least match the processing capabilities
possible for light modes.
In Sec.~\ref{enta} we show that,
even restricting to cases where only global control of the trapping potentials is possible,
such systems are actually apt to outperform optical modes in the generation of entanglement,
both bipartite and multipartite.
As applications we consider,
in Sec.~\ref{propa}, the
propagation of quantum information along the array of ions,
at both qubit and continuous variable levels
and, in Sec.~\ref{bell}, the violation of multipartite
non locality tests, and show them to be within the reach of current technology.
Radial modes will thereby turn out to be promising not only for information processing but also
as probes of fundamental physics.

\begin{figure}[t!]
\begin{center}
\includegraphics[scale=0.4]{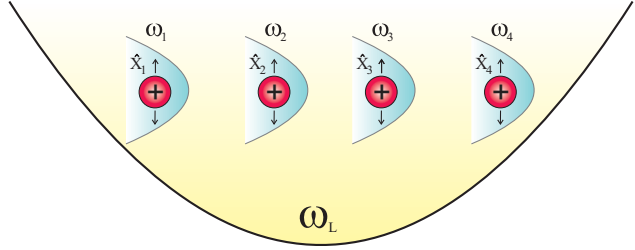}
\caption{Considered setup: the ions are trapped along the longitudinal direction
with the same trapping frequency $\omega_{L}$. Also, each ion $j$ is trapped
along the transverse, `radial' direction ({\em i.e.}, along the arrows in the drawing)
with trapping frequency $\omega_{j}$.
The operator $\hat{X}_{j}$ stands for the position of ion $j$ along the radial direction:
these radial oscillations are described by the modes $\{\hat{X}_{j},\hat{P}_{j}\}$ we shall focus on.
\label{setting}}
\end{center}
\end{figure}
\section{The trap}\label{trap}

We shall consider the radial modes
of $n$ ions of mass $m$ and charge $ze$ in a linear Paul trap
\cite{james98}.
Let $\hat{X}_{j}$ and $\hat{P}_{j}$ be the
position and momentum operators associated to the radial degree
of freedom of the $j$-th ion, which is trapped in the radial
direction with angular frequency ${\omega}_{j}$ (see Fig~\ref{setting}).
In the following, the {\em longitudinal}
trapping frequency $\omega_L$ will be the unit of frequency and will
set the unit of length as well (equal to
$d=\sqrt[3]{z^2e^2/(4\pi\varepsilon_{0}m\omega_{L}^2)}$, where
$\varepsilon_0$ is the dielectric constant); also, we shall set
$\hbar=1$, so that all the quantities will be dimensionless.
The Coulomb interaction affects the local radial
oscillation frequencies: for convenience, let us then define the
`effective' local radial frequencies
$$\tilde{\omega}_{j} =
\omega_L\sqrt{\frac{{\omega}^2_j}{\omega_L^2}-\sum_{l\neq j}
\frac{d^3}{|u_{j}-u_{l}|^{3}}} \; ,$$
$\{u_j\}$ being the equilibrium positions of the ions \cite{james98}.
Rescaling the canonical operators according to
$\hat{x}_{j}= \sqrt{m\tilde{\omega}_j}\hat{X}_{j}$,
$\hat{p}_{j}= \hat{P}_j/\sqrt{m\tilde{\omega}_j}$, and grouping them
in a vector of operators
$\hat{R}=(\hat{x}_1,\ldots,\hat{x}_{n},
\hat{p}_j,\ldots,\hat{p}_{n})^{\sf T}$, allows one to express
the global Hamiltonian of the system
in the harmonic approximation\footnote[1]{Notice that,
in the harmonic approximation ({\em i.e.}, at second order in the
ions' displacements) the coupling between radial modes and longitudinal ones vanishes.
Fourth order couplings are around $10^{-4}$ times smaller in the considered
experimental conditions and will thus be safely neglected.}
as
\be \label{hamil}
\hat{H} = \frac12 \hat{R}^{\sf T}
\left(\gr{\kappa}\oplus
\gr{\tilde{\omega}}\right)
\hat{R} \; ,
\ee
where $\gr{\tilde{\omega}}$ is a diagonal matrix:
$\gr{\tilde{\omega}}={\rm diag}\,(\tilde{\omega}_1,\ldots,\tilde{\omega}_{n})$,
while the potential matrix $\gr{\kappa}$ has diagonal entries
$\kappa_{jj}=\tilde{\omega}_j$ and off-diagonal entries
$\kappa_{jk}=\omega_L^2 d^3/(\sqrt{\tilde{\omega}_{j}\tilde{\omega}_{k}}|u_j-u_k|^3)$
for $j\neq k$.

Let us recall the canonical commutation \label{uus}
relations
$[\hat{R}_j,\hat{R}_k]=i\Omega_{jk}\id$, where the $2n\times2n$
matrix
$\Omega$ has entries $\Omega_{j,k}=\delta_{n,k-j}-\delta_{n,j-k}$
for $1\le j,k\le2n$, and that
Gaussian states are defined as states with Gaussian characteristic
functions: a Gaussian state $\varrho$ is thus uniquely determined
by its ``covariance matrix'' (CM) ${\gr\sigma}$, with entries given by
\be
{\sigma}_{jk} \equiv \frac{1}{2}\tr{[\{\hat R_j , \hat R_k\} \varrho]}
-\tr{[\hat R_j \varrho]}\tr{[\hat R_k \varrho]} \label{cm} \; ,
\ee
and by the vector
of first moments $R$, with components $R_{j}\equiv \tr{[\hat{R}_j \varrho]}$
\cite{martinrev,martinshash}.

The ground state of Hamiltonian $\hat{H}$
is a Gaussian state with a block diagonal CM
$\sig_{g}=(\sig_{x}\oplus\sig_{x}^{-1})/2$, where
$$\sig_{x}=\gr{\tilde{\omega}}^{1/2}(\gr{\tilde{\omega}}^{1/2}\gr{\kappa}\gr{\tilde{\omega}}^{1/2})^{-1/2}\gr{\tilde{\omega}}^{1/2} \; ,$$
and vanishing first moments.

Finally, let us remember that the evolution for the time $t$ of an initial Gaussian
state with CM $\sig$ under a quadratic Hamiltonian $\hat{H} =
\hat{R}^{\sf T} H \hat{R}$ (where $H$ is any symmetric matrix)
is a Gaussian state with CM given by \be \sig_{t} = S_{t}\sig
S_t^{\sf T}\; ,  \label{harmevo} \ee for $S_{t} = \exp(\Omega Ht)$.

\subsection{Dissipation}

In our study, we will take into account the decoherence of the radial
modes in an environment of phonons with temperature $T$ and `loss rate'
$\gamma$ (for simplicity assumed to be the same for every mode).
Under such conditions, the evolution of the ions' state $\varrho$ at frequencies $\{\tilde{\omega}_j\}$
is described by the following master equation in interaction picture
\cite{serafozzi05}
\be
\hspace*{-2.3cm}
\frac{{\rm d}\varrho}{{\rm d}t} = \frac{\gamma}{2} \sum_{j=1}^{n}\left[
N_j (2a_j^\dag \varrho a_j - a_j a_j^\dag \varrho - \varrho a_j a_j^{\dag} )
+ (N_j+1) (2a_j \varrho a_j^\dag -a_j^\dag a_j \varrho - \varrho  a_j^{\dag} a_j)
\right] , \label{master}
\ee
where the number of phonons in the radial mode $j$ is given by
$N_j := 1/\left({{\exp}\left(\frac{\hbar \tilde{\omega}_j}{k_B T}\right)+1}
\right)$,
according to Bose law ($k_B$ being Boltzmann constant) and $a_{j}:=
(\hat{X}_j+i\hat{P}_j)/\sqrt{2}$.
In accordance with the experimental terminology,
we will also refer to the quantity $\epsilon_j := N_j\gamma$ as the
`heating rate', essentially
representing the rate at which thermal phonons are injected into mode $j$.

Eq.~\ref{master} preserves the Gaussian character of the initial state and 
can results, for the CM $\sig$ of Gaussian states, into the following equation: 
$$
\frac{\rm d}{{\rm d}t} \gr \sigma = \gamma \sig_\infty - \gamma \sig 
$$
(where $\sig_{\infty}\equiv \sig'\oplus\sig'$,
$\sig' = {\rm diag}(N_1+\frac12,\ldots,N_n+\frac12)$), with solution:
$${\rm e}^{-\gamma t}\sig_{t}+(1-{\rm e}^{-\gamma t})
\sig_{\infty} \, .$$

\section{Linear phononics and beyond}\label{manip}

In this section, we shall assume that the trapping frequencies $\{\omega_j\}$,
and thus $\{\tilde{\omega}_j\}$, can be controlled locally and changed
suddenly ({\em i.e.}, much faster than $\omega_{j}^{-1}$).
Such a local control may be achieved by building small, local radial electrodes, by
adding local optical standing waves,\footnote[2]{Currently, standing waves realize, at most, trapping potentials
of about $1\,{\rm MHz}$ in experiments where the internal degrees of freedom are controlled.
However, the manipulation of radial modes can tolerate
much higher scattering rates than that of internal degrees of freedom,
and would thus allow for much higher trapping frequencies.}
or in
trap arrays \cite{stahl2005,Ciaramicoli2005}.
Let us also note that a setting formally identical to the one we describe here could be reproduced shortly 
even for locally defined {\em axial} (`longitudinal') modes in segmented Paul traps \cite{schmidtkaler}. 
The changes in trapping frequency we will consider (typically ranging between $1$ and $10$ ${\rm MHz}$)
can be realised in about $10 {\rm ns}$ \cite{labaziewicz},
and will be regarded as instantaneous over our timescales.
Our aim is
to show how, in principle, local control of the frequencies allows one to perform any
arbitrary ``linear optical'' operation on the radial modes of the
ions, that is any unitary operation under which, in the Heisenberg
picture, the vector of operators $\hat{R}$ transforms linearly:
$\hat{R}\mapsto S \hat{R}$. The matrix $S$ has to be `symplectic',
{\em i.e.} such that
$$
S^{\sf T}\Omega S=\Omega
$$
(where $\Omega$ is the anti-symmetric symplectic form, defined
at page \pageref{uus}), to preserve the canonical commutation
relations. Notice that such a class of operators includes squeezing
transformations.\footnote[3]{In the literature, squeezing
transformations have a tendency to be excluded from the class
of ``linear'' transformations, because their implementations
require `nonlinear' (third order) interactions. However, the
effective evolution of the modes of interest (`signal' and
`idle', usually) is actually linear in the sense specified
above (linear evolution of the vector $\hat{R}$ in Heisenberg
picture), and we will thus generally refer to `linear'
transformations as including squeezing.} We should mention
that the idea of squeezing the state of a single ion by
controlling the trapping frequency was first put forward in
the early 90's \cite{Heizen1990}. Also, single-mode squeezing
can be achieved by coupling the motion of the ion with its
internal degrees of freedom \cite{cirac1993s,solano05}. The
scope of the present Section is however wider, since we
set out to prove that any symplectic operation on any number
of ions can be implemented by frequency manipulation.

Let us first remark that any symplectic operation $S$ on a system of
many canonical degrees of freedom (``modes'') can be decomposed into
a combination of generic single-mode symplectic transformations and
two-mode rotations (``beam splitters'', in the quantum optical
terminology) \cite{pramana,reck94}.
This fact follows from the Euler decomposition of symplectic operations \cite{pramana}
and from the possibility of decomposing energy preserving operations into a network of beam-splitters
between pairs of modes \cite{reck94}.
It is therefore sufficient for
us to establish the possibility of performing these two subclasses of
operations (single-mode symplectic transformations and two-mode beam splitters)
on our system of $n$ ions by manipulating the local frequencies.
In turn, again because of Euler decomposition,
single-mode operations can always be reduced to combinations of
squeezings and phase-shifts.

\subsection{Single mode operations}
In what follows we
assume that the original effective frequencies of the ions are different but
commensurate, as given by, say, $\tilde{\omega}_{j} = j \tilde{\omega}$, and that $\tilde{\omega}$ is
large enough so that interaction between any two ions is
suppressed when the relevant coherent manipulation sets off.\footnote[4]{The linear scaling of the coupling is not a stringent requirement.
Because the coupling between ion $j$ and $k$ falls off like $|j-k|^{-3}$, one can safely assume,
say, a `triangle' profile (alternately increasing and decreasing) for the couplings in long chains of ions.
Note also that, for a linear scaling, the population of the levels of ion $j$ due to the interaction with $k$ roughly scales as
$1/(1+(\tilde{\omega}/\omega_L)^2|j-k|^5)$: a frequency step $\tilde{\omega}\simeq 20 \omega_L$ is already enough
to make the effects of all the interactions essentially negligible.}
Let us then consider the reaction of the system if the frequency
of the $j$-th ion changes suddenly from $\tilde{\omega}_{j}$ to
$\alpha_{j}\tilde{\omega}_{j}$, for some real $\alpha_{j}$
(clearly, such changes are completely governed by changes in the $\omega_{j}$'s).
The Heisenberg equation of motion for $\hat{x}_j$ and $\hat{p}_j$
can be immediately integrated in such a case, resulting into a
symplectic transformation $S_j(t)$ given by
\be
\hspace*{-1.65cm}S_j(t) = \left(\begin{array}{cc}
\alpha_j^{\frac12} & 0 \\
0 & \alpha_j^{-\frac12}
\end{array}\right)
\left(\begin{array}{cc}
\cos(\tilde{\omega}_j\alpha_j t) & \sin(\tilde{\omega}_j\alpha_j t) \\
-\sin(\tilde{\omega}_j\alpha_j t) & \cos(\tilde{\omega}_j\alpha_j t)
\end{array}\right)
\left(\begin{array}{cc}
\alpha_j^{-\frac12} & 0 \\
0 & \alpha_j^{\frac12}
\end{array}\right)
\label{simple} \, .
\ee
The first and last matrix of this decomposition correspond to
`squeezing' operations in the quantum optical terminology,
whereas the second factor is known as a `phase shift'
({\em i.e.}, a rotation in the single-mode phase space).
Combinations of squeezings and phase-shifts make up any
possible single-mode symplectic operation: we thus need
to show that such operations can be implemented individually
on any ion of the system in a controllable manner
(as pointed out above, this is sufficient
because of the Euler decomposition of single mode operations \cite{pramana}).

\subsubsection{Phase-shift}\label{phaseshift}
To realise a phase-shift operation on the $k$-th ion,
without any squeezing,
it is sufficient to change
the frequencies of all the other ions in the same way,
such that, in the notation defined above, $\alpha_k=1$ and $\alpha_j=\alpha\neq 1$ for
$j\neq k$ ($\alpha_{j}$ being the factor by which the frequency of ion $j$ is
multiplied by manipulating the local trapping potential).
As apparent from \eq{simple}, after a time
$t_{\alpha}=2\pi/(\tilde{\omega}\alpha)$ from the change one has $S_{j}=\id_{2}$
for $j\neq k$: all the ions which undergo a change
in trapping frequency are back to their initial state
(let us recall that $\tilde{\omega}_j=j\tilde{\omega}$ by assumption, so that after $t_{\alpha}$
the central rotation of \eq{simple} reduces to the identity and the two
opposite squeezing transformations just nullify each other).
On the other hand, the transformation $S_{k}$ will be equal to
the central rotation (with no squeezing,
as $\alpha_k$ is kept equal to $1$) of angle $\varphi_k=2\pi k/\alpha$:
the oscillation of the $k$-th ion will have acquired such a
phase, analogous to the ``optical phase'' of light modes.
If the frequencies are switched back to the original values after a time
$t_{\alpha}$, the net effect of the evolution is then
analogous to an `optical' phase-shift of angle $\varphi_k$ on the ion $k$.

\subsubsection{Squeezing}\label{squeeze}
In order to squeeze the state of ion $k$, leaving all the other ions
unaffected, one can conversely
change only the pertinent frequency, so that $\alpha_k\neq 1$
and $\alpha_j=1$ for $j\neq k$. Then, after a time period
$t=2\pi/\tilde{\omega}$, all the other ions will have returned to the
initial state, while ion $k$ will be squeezed and phase-shifted
according to \eq{simple}.
The phase-shift can always
be arbitrarily corrected by applying the strategy to obtain ``pure'' phase shifts
described in Sec.~\ref{phaseshift}, which conclusively shows that
{\em any linear operation on a single mode can be implemented by controlling
the local trapping frequencies}.\footnote[5]{In fact, arbitrary adjustment of the phase
allows for ``pure'' squeezing operations along the
phase space directions $x$ and $p$, and thus implies the possibility to construct any
linear operation, because of the symplectic Euler decomposition previously discussed.}

The degree of squeezing achievable with such a strategy
depends crucially on the phase-shift operation since,
as shown by \eq{simple},
the two squeezing operations act along orthogonal directions and are
the inverse of each other. In the case $\alpha_{k}=(\frac14 +
h)/k$ for $h\in {\mathbbm{N}}$,
the two squeezings act effectively
along the same phase space direction: in this instance
the phase-shift can be balanced by a counter-rotation of $\pi/4$ in
phase space and the final squeezing operation is a diagonal matrix
given by ${\rm diag}(\alpha_k,\alpha_k^{-1})$.
In principle, the value of $\alpha_{k}$ is only
limited by the stability of the system and
the breakdown of the harmonic approximation \cite{nonlinear},
occurring when the squeezing is comparable to the
ratio between the size of the wave packet and the distance between the atoms. In actual experiments,
the considered setup would thus allow for
$\sqrt{\alpha_{k}} \ll 500$.
Notice that such values are by far out of reach in optical systems,
where squeezings corresponding to $\alpha_k\simeq 8$ were recently reported
\cite{takeno07}.
Also notice that, by placing the ions inside cavities, the squeezing of
the massive degrees of freedom could be transferred to light, so that
radial modes could act as an effective source of squeezing (and
potentially even entanglement) for optical systems as well \cite{noialtri}.

\subsubsection{Beam-splitters}\label{bs} To extend our proof to any linear operation,
on any number of ions, let us now turn to `beam-splitting'
operations between the radial modes of any two ions in the array.
Such operations are achieved by bringing the two modes
in question (hereafter labeled by $j$ and $k$) to the same frequency
$\tilde{\omega}=\tilde{\omega}_j=\tilde{\omega}_k$, so that the
Coulomb interaction between them is no longer suppressed.
Switching to the interaction picture, one has the following coupling
Hamiltonian between the two modes:
$$
\hat{H}_{I} =
\kappa_{jk}\hat{x}_j(t)\hat{x}_k(t) = \kappa_{jk} (a_{j}\,{\rm
e}^{-i\tilde{\omega} t}+a_{j}^{\dag}\,{\rm e}^{i\tilde{\omega} t}) (a_{k}\,{\rm e}^{-i\tilde{\omega}
t}+a_{k}^{\dag}\,{\rm e}^{i\tilde{\omega} t}) \, ,$$
where the ladder operators are
defined as $\hat{x}_j=(a_j+a^{\dag}_j)$ and $\kappa_{jk}$ is an entry of the matrix
$\gr{\kappa}$ defined in \eq{hamil}.
If the frequency $\tilde{\omega}$ is
sufficiently large
the rotating wave approximation reliably applies and the rotating terms can be
neglected to yield\footnote[6]{This always holds
at the trapping frequencies we shall consider in specific examples,
going from $1$ to few tens of ${\rm MHz}$, for which stability around $10^{-3}$ are
achievable.}
$$ \hat{H}_{I} =
\kappa_{jk} (a_{j}a_{k}^{\dag}+a_{j}^{\dag}a_{k}) \;.$$
This Hamiltonian realises exactly the desired beam splitter-like
evolution, resulting into a symplectic transformation which mixes
$\hat{x}_j$ with $\hat{x}_k$ and $\hat{p}_j$ with $\hat{p}_k$
(rotating such pairs equally, by the angle $\kappa_{jk}t$). For
instance, a `$50:50$' beam splitter is achieved after a time
$t=\pi/(4\kappa_{jk})$. Since the interaction requires a change of
the local frequencies it includes automatically in it a local
operation, which may however be corrected before or after the
`beam-splitting' procedure, as detailed above.

Summing up, the combined arguments presented in Secs.~\ref{phaseshift}, \ref{squeeze} and \ref{bs}
show that {\em any ``linear optical'' operation, {\em including
squeezing operations and entangling operations acting on multiple different modes},
can be implemented for radial modes of trapped ions by properly tuning the
frequencies of the radial microtraps of the individual ions.}

Linear optical
operations, complemented by displacements (see Sec.~\ref{further}), correspond
to all the unitary transformations that preserve the Gaussian character of the initial state.
Therefore, all the developments based on Gaussian states in quantum optics,
in particular concerning entanglement manipulation
\cite{martinrev} and information protocols \cite{braunstein05},
can be carried over to radial modes of ion traps if local control is achieved.
Arguably, the framework we have outlined for the implementation of linear optical operations
on the local oscillations of distinct ions, to which one may refer as ``linear phononics'',
could offer further practical advantages over its quantum optical counterparts,
essentially related to the static nature of the quantum information processed
(for instance, `mode matching', which hampers substantially the implementation of linear optics,
is not a problem here).


\subsection{Further manipulations and measurements}\label{further}

Beside the linear operations treated so far,
implemented through varying trapping potentials and Coulomb interactions,
the motional degrees of freedom of trapped ions allow for other controlled manipulations,
some of which take advantage of the coupling between motion and the internal degrees of freedom
that can be engineered by applying standing- or traveling-wave pulses to the ions.
In order to give a complete account of the possibilities offered by radial modes, let us here review
such strategies, and briefly comment about their applicability to
radial modes. As a general remark let us mention that, because of
the tighter confinement they allow for, radial modes easily meet
the Lamb-Dicke condition (depending on the width of the ground
state's wavepacket), which means that the coupling with the internal
degrees of freedom can be tailored to a high degree of accuracy
(generally better than for longitudinal modes).\footnote[7]{The
price to pay for such an accuracy is longer operation times, which
could be ultimately reduced if stronger lasers became viable.}
Also, individual ions can be addressed in such manipulations as, at spacings of some micrometers
and assuming pulses' waists of the order of ${\rm 1}$ $\mu{\rm m}$, less than ${\rm 1}\%$ of
the central laser power would be shined on neighbouring ions with respect to the central one.

{\em Displacement} operations, which shift the
operators $\hat{R}_j$ by real numbers, can be
realized in several ways: by classical driving fields,
by standing waves, or by shifting the radial equilibrium positions
of the ions \cite {Meekhof1996}.
In particular, transverse driving fields could be applied to displace the radial
modes, with no particular hurdles.
In the following, the unitary displacing the canonical operators of mode $j$ by,
respectively, $x_{j}$ and $p_{j}$ will be denoted by
$\hat{D}_j(x_j,p_j)$.

Notably, even {\em non-Gaussian} states can be engineered
in this setup with relative ease (with respect to quantum optics, where
non-Gaussian manipulations require higher order nonlinearities, usually extremely weak),
either by entering the nonlinear regime of the Coulomb interactions or by
coupling the internal degrees of freedom of the ions.
The experimental realisation of a cross-Kerr coupling for the longitudinal oscillations of
trapped ions has been recently reported \cite{nonlinear}.

The coupling to internal degrees of freedom
also allows for Gaussian and non Gaussian measurements on individual ions.
The measurement of quadrature operators, corresponding to {\em homodyning},
was proposed in \cite{Vogel1995,Poyatos1996a,Bardroff1996}.
Quantum non-demolition measurement of local number states and parity could be measured as well, by applying the scheme suggested and realised for cavity QED in \cite{Gleyzes2007}, based on the dispersive coupling of the
number of oscillations to two internal levels of the ion. Quite
remarkably, such a scheme would allow one to measure the phonon
number's parity on a single copy of the state and run of the
apparatus. As we will see in Sec.~\ref{bell}, this possibility
is consequential for the violation of Bell-like inequalities.
Notice also that such non-Gaussian measurements pose a considerable
technological challenge for light fields, where resolving photon
numbers with high detector efficiency is still daunting despite
recent progress \cite{walmsley,vlpc}. On the other hand,
the schemes recalled above are bound to be comparatively slow,
requiring to wait for half a period of coherent interaction between
motions and internal levels and then subsequent readout of the
internal levels by, {\em e.g.}, fluorescence. When internal degrees
of freedom are involved, radial modes require longer times, roughly
on the order of tens of microseconds, but achieve remarkable precision.

Finally, concerning the preparation of initial states for coherent
manipulations, let us remind that cooling of ion oscillations
to their ground state can be achieved very efficiently by
sideband-Raman pulses \cite{winelandold,monroe1995} (or more
recent variations over such a strategy
\cite{MorigiKeitel,alexmartincool}). Note, since in the described scenario the initial local potential is much larger than the interaction, simultaneous cooling of all the chain could be done by cooling each ion to its local ground state, which is to a good approximation the global ground state.
The local single excitations on such ground states
may be prepared by addressing the first ion with a proper sequence of blue and red sideband pulses, as detailed in \cite{Meekhof1996}.

In the remainder of the paper we shall demonstrate the
potential of linear operations on radial modes in specific applications,
with a particular focus on settings fully
accessible to current experiments.
\begin{figure}[t!]
\begin{center}
\begin{tabular}{cc}
\includegraphics[width=6cm,height=5.5cm]{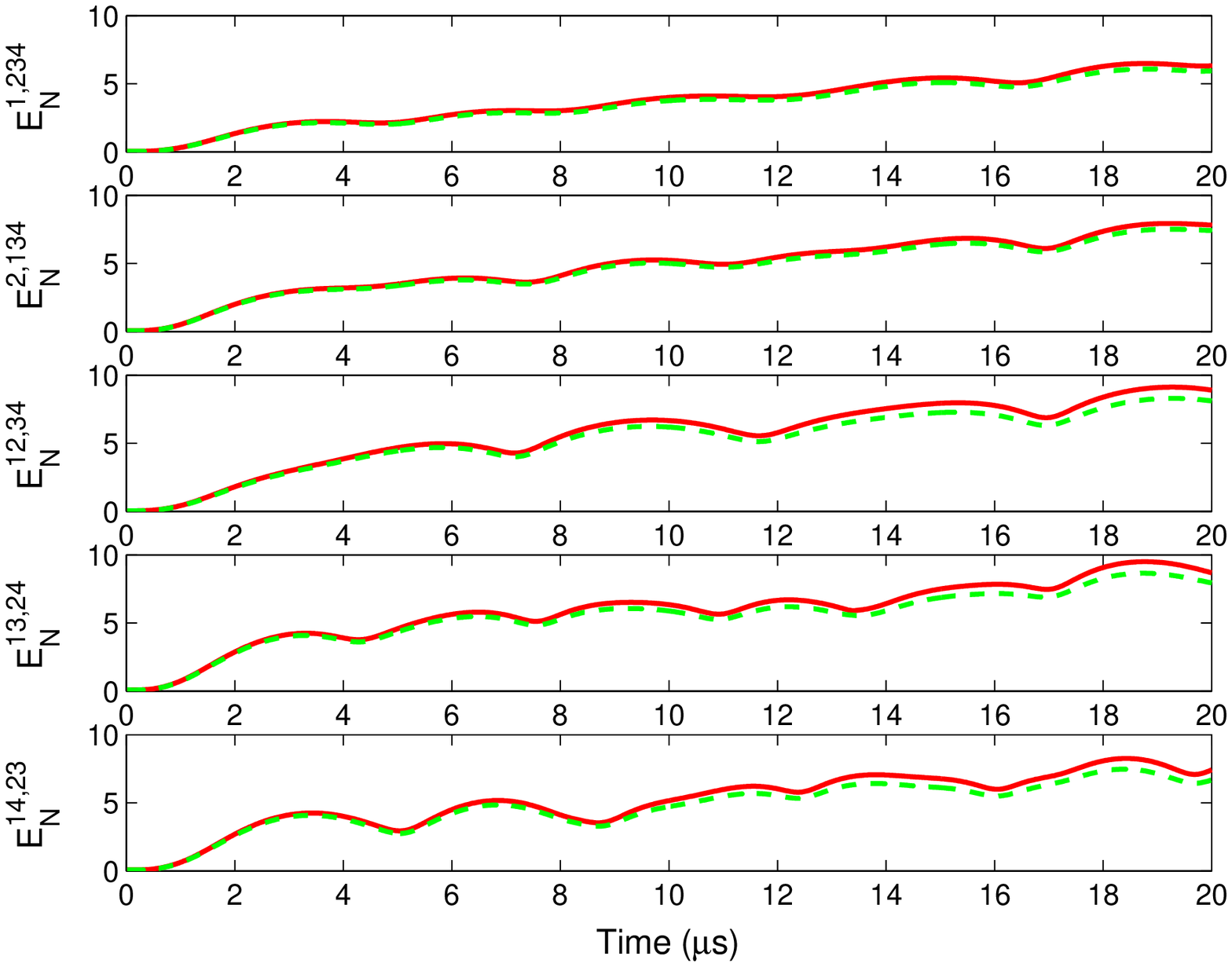}&
\includegraphics[width=6cm,height=5.5cm]{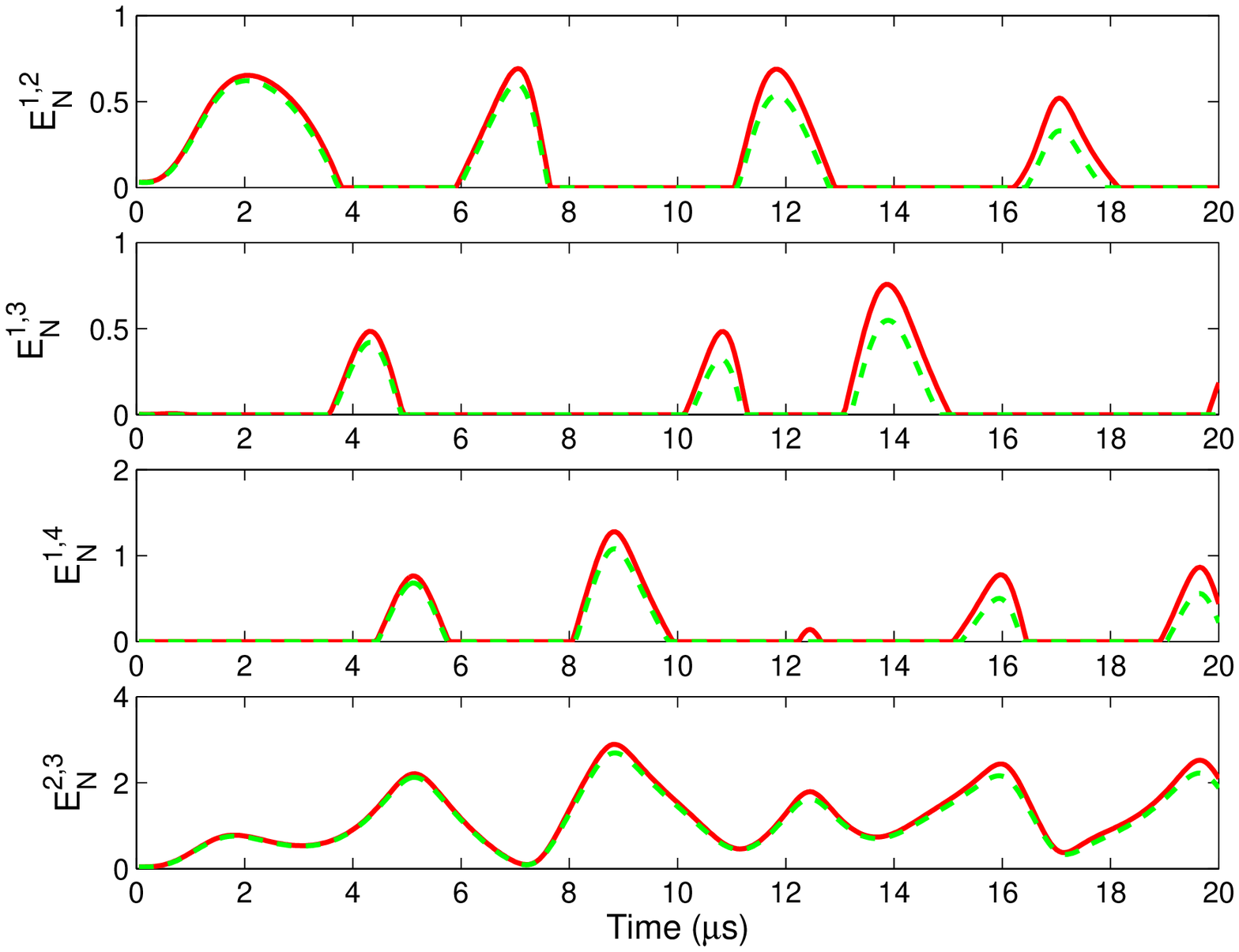}
\end{tabular}
\caption{Dynamics of the entanglement (in ebits of logartithmic negativity)
between the radial modes of the ions, in a trap
containing $n=4$ ions and with longitudinal trapping frequency $\omega_{L}=1{\rm MHz}$.
At time $t=0$ the system starts off from the ground state for transverse trapping frequency
$\omega_{i}=5{\rm MHz}$ and then evolves under a trapping frequency $\omega_{f}=2{\rm MHz}$.
The plots on the left show the logarithmic negativity shared by all the possible bipartitions
of the four modes (because of the spatial symmetry, one has only five distinct cases in the chain:
for instance, one has $E_{\N}^{1,234}=E^{4,123}_{\N}$). On the right, the logarithmic negativity
shared by pairs of modes is plotted (again, because of symmetry, there occur only four distinct cases).
The continuous (red) curves refer to a case with no dissipation, while the dashed (green) curves
refer to a case with heating rate $\epsilon=2{\rm KHz}$ (resulting from a coupling to the bath
$\gamma=10^{-4}{\rm Hz}$ and a temperature $T=294^{\circ}{\rm K}$).
\label{multienta4}}
\end{center}
\end{figure}
\section{Entanglement generation}\label{enta}

The set of operations described in Sec.~\ref{manip}, including local
squeezing and two-mode beam splitters between distant modes, allows
for the generation of entanglement between the radial modes of two
ions in the chain. In fact, as is well known, two single-mode states
squeezed along orthogonal quadratures, entering a beam splitter, give
rise to an entangled state for the outgoing modes (this is a standard
procedure to create CV entanglement in quantum optics).
In principle, for the system of trapped ions under examination,
the local control of the transverse trapping frequencies allows for the creation of tailored entangled states,
where the entanglement can be build up between any two ions in the chain.
Moreover, and possibly even more intriguingly, the possibility of controlling a system of $n$ parties (the ions)
all constantly interacting with each other, paves the way for the creation of {\em multipartite} entanglement
between the distinct ions.

In this section we show that
unprecedented degrees of bipartite entanglement, as well as
interesting and robust multipartite entangled states, can be obtained
in settings accessible to current experiments,
starting from a ground state of the system and {\em requiring only global control
of the trapping potential} (where, {\em i.e.}, the trapping frequencies $\{\omega_{j}\}$
are the same for all ions at any time, and are only changed simultaneously).
Further, we will briefly discuss the possibilities
a local control of the frequencies that would open up for the
controlled dynamics of multipartite entanglement.

The specific situation we shall address starts off from the ground state
$\varrho_{g}$ of Hamiltonian $\hat{H}$ [see \eq{hamil}] -- with all
frequencies being equal, {\em i.e.~}$\omega_{j}=\omega_i$
for $1\le j\le n$ -- as the initial state.
Next, the trapping frequencies are changed to the common value $\omega_f$,
so that the state $\varrho_{g}$ will not be
stationary anymore under the modified Hamiltonian.
For large
$\omega_i$, the initial state $\varrho_{g}$ contains very little
entanglement but, if the frequencies are suddenly changed, entanglement builds up during the subsequent
evolution (see \cite{nanos} for an analogous scheme in chains
of nanomechanical oscillators).
Entanglement between different subsystems will be quantified by the
logarithmic negativity $E_{\N}\equiv \log_{2}\|\tilde{\varrho}\|_1$,
where $\|\tilde{\varrho}\|_1$ stands for the trace norm of the
`partially transposed' density matrix of the subsystem \cite{gaussneg,martinshash}.
Different partitions will be denoted by superscripts referring to the ions:
for instance, the logarithmic negativity between ions $1$-$2$ and $3$-$4$ will be
denoted by $E_{\N}^{12,34}$.
{The intuition behind this entanglement generation method is that
operating faster than the speed of sound in the system,
{\em i.e.}, essentially, faster than the inverse of the energy gaps,
is analogous to operating locally.
Thus, by changing the local potentials (in the same way for each ion)
fast enough one generates local squeezing,
which is then `converted' into entanglement by the
time-evolution through the harmonic coupling between the ions,
analogous to a set of beam splitters.}

\begin{figure}[t!]
\begin{center}
\begin{tabular}{cc}
\includegraphics[width=6cm,height=5cm]{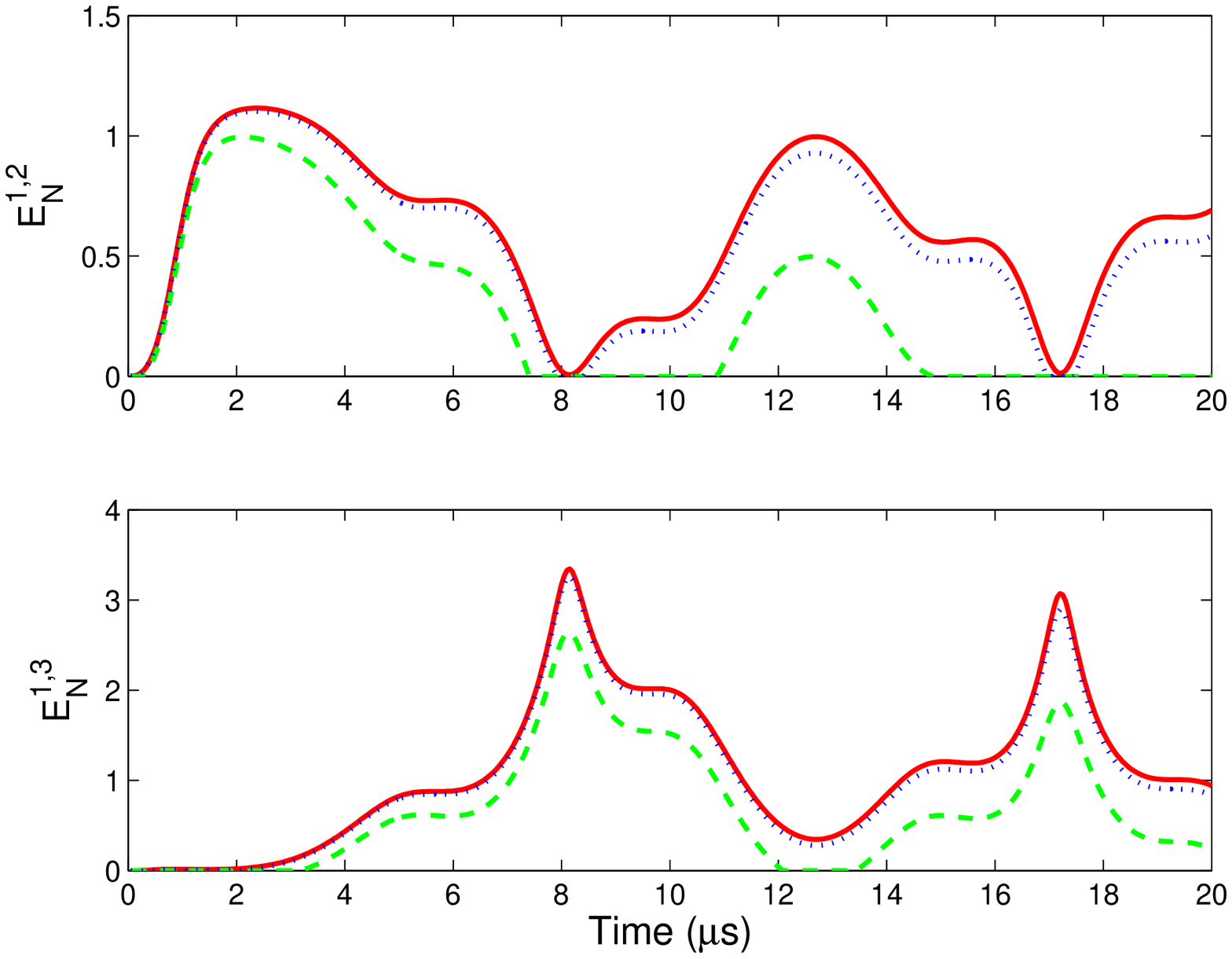}&
\includegraphics[width=6cm,height=5cm]{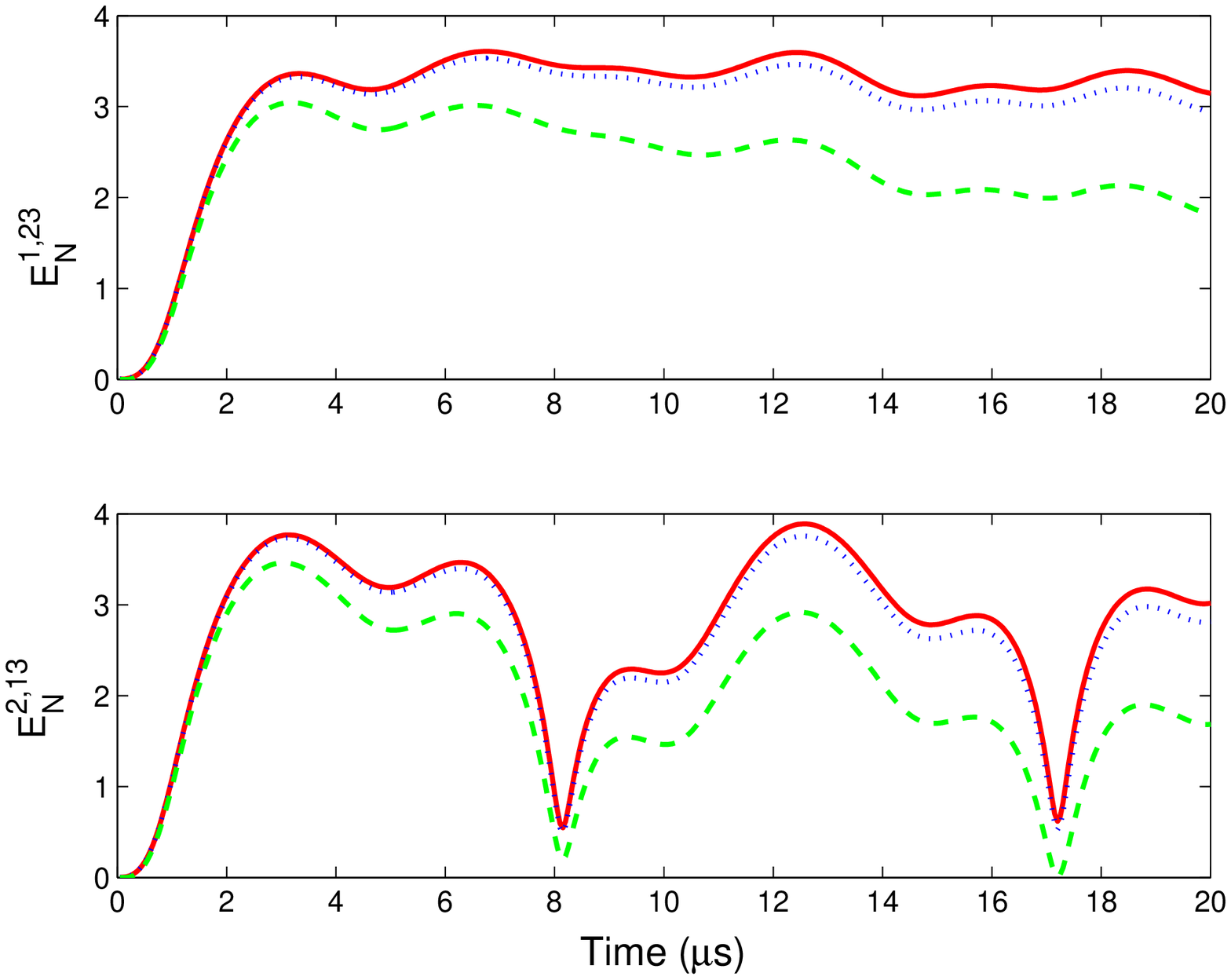}
\end{tabular}
\caption{Dynamics of the entanglement (in ebits of logarithmic negativity)
between the radial modes of the ions, in a trap
containing $n=3$ ions and with longitudinal trapping frequency $\omega_{L}=1{\rm MHz}$.
At time $t=0$ the system starts off from the ground state for transverse trapping frequency
$\omega_{i}=20{\rm MHz}$ and then evolves under a trapping frequency $\omega_{f}=2{\rm MHz}$.
The plots on the left show the logarithmic negativity shared by all the possible bipartitions
of the four modes (because of the spatial symmetry, one has only two distinct cases in the chain).
On the right, the logarithmic negativity
shared by pairs of modes is plotted (again, because of symmetry, there occur only two distinct cases).
The continuous (red) curves refer to a case with no dissipation,
the dotted (blue) curves refer to a case with heating rate
$\epsilon=200{\rm Hz}$ (resulting from a coupling to the bath
$\gamma=10^{-5}{\rm Hz}$ and a temperature $T=294^{\circ}{\rm K}$),
while the dashed (green) curves
refer to a case with heating rate $\epsilon=2{\rm KHz}$ (resulting from a coupling to the bath
$\gamma=10^{-4}{\rm Hz}$ and a temperature $T=294^{\circ}{\rm K}$).\label{multienta3}}
\end{center}
\end{figure}

The case portrayed in Fig.~\ref{multienta4} represents four ions
starting from the ground state
for $\omega_i=5{\rm MHz}$, which then evolves under the frequency
$\omega_{f}=2{\rm MHz}$ for $20 \mu{\rm s}$ (for a transverse trapping frequency $\omega_{L}=1{\rm MHz}$).
As can be seen, the global state gets entangled under any possible
bipartition of the four modes, a situation which is referred to in the literature as
``complete inseparability'' \cite{giedke01}.
A more extensive inspection shows that this is a general property of the Hamiltonian at hand, even for larger
number of ions.
Of course, as shown in Fig.~\ref{multienta3}, complete inseparability also occurs for
three ions: in Sec.~\ref{bell} we shall see how the multipartite
entanglement for three ions analysed in Fig.~\ref{multienta3} could serve to violate Bell-like inequalities.
Our plots also show that such an entanglement exhibits considerable resilience under realistic heating rates.
In this respect,
notice that the heating rate for the dashed (green) curved plotted in the graphs is $\epsilon=2 {\rm KHz}$,
and that heating rates as low as $\simeq10{\rm Hz}$ have been recently observed
in ion traps (even though for a single ion only \cite{Garg,Deslauriers}):
robust multipartite entanglement (between
$5$ and $10\, {\rm ebits}$ of logarithmic negativity) between the ions can thus be created.
Such degrees of
entanglement are inconceivable for multipartite systems in photonic systems
where, furthermore, the manipulation of several modes tends to become awkward due
to the increasing number of mode-matching conditions to be fulfilled.

\begin{figure}[t!]
\begin{center}
\subfigure[a]
{\includegraphics[width=6cm,height=5cm]{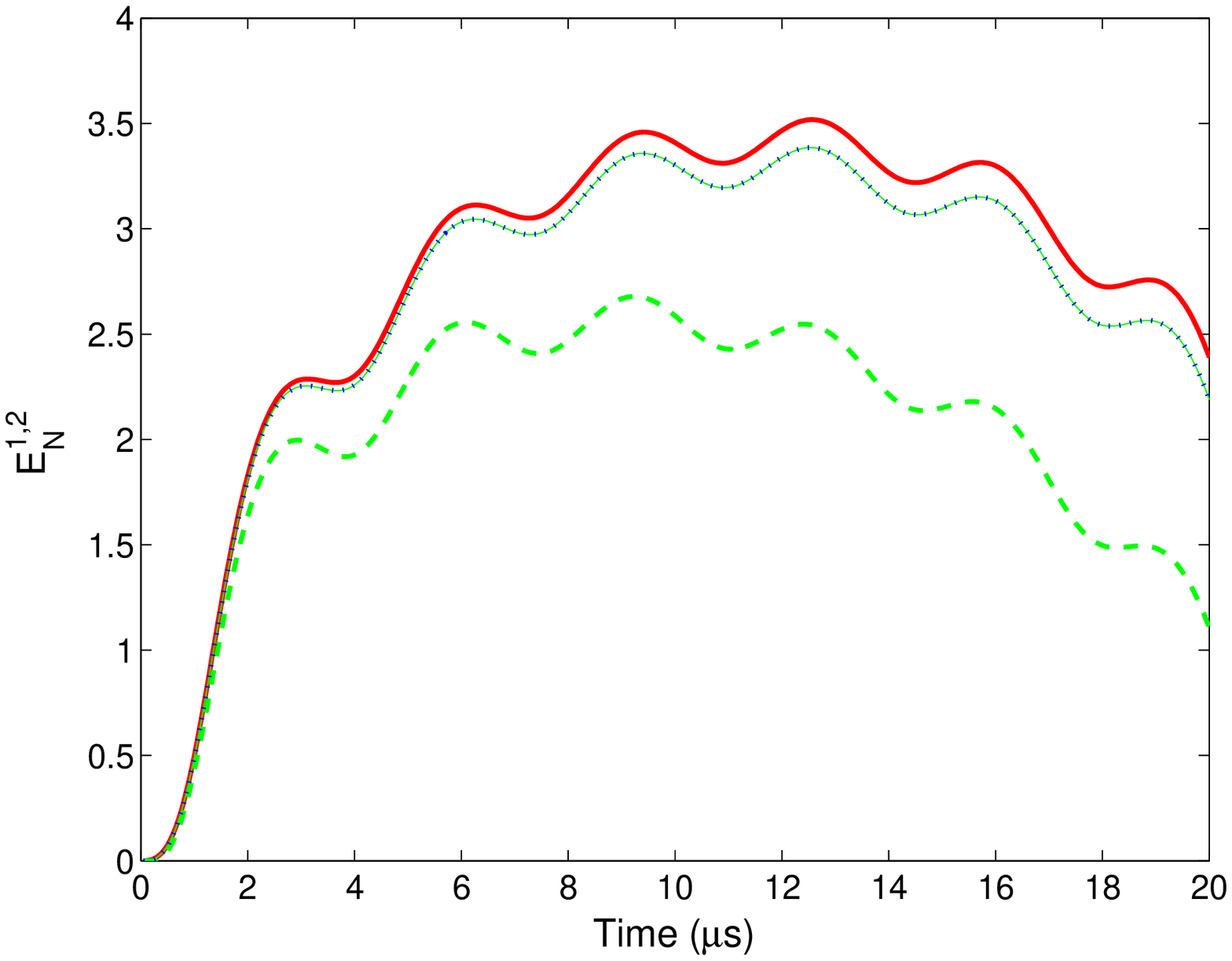}}
\subfigure[b]
{\includegraphics[width=6cm,height=5cm]{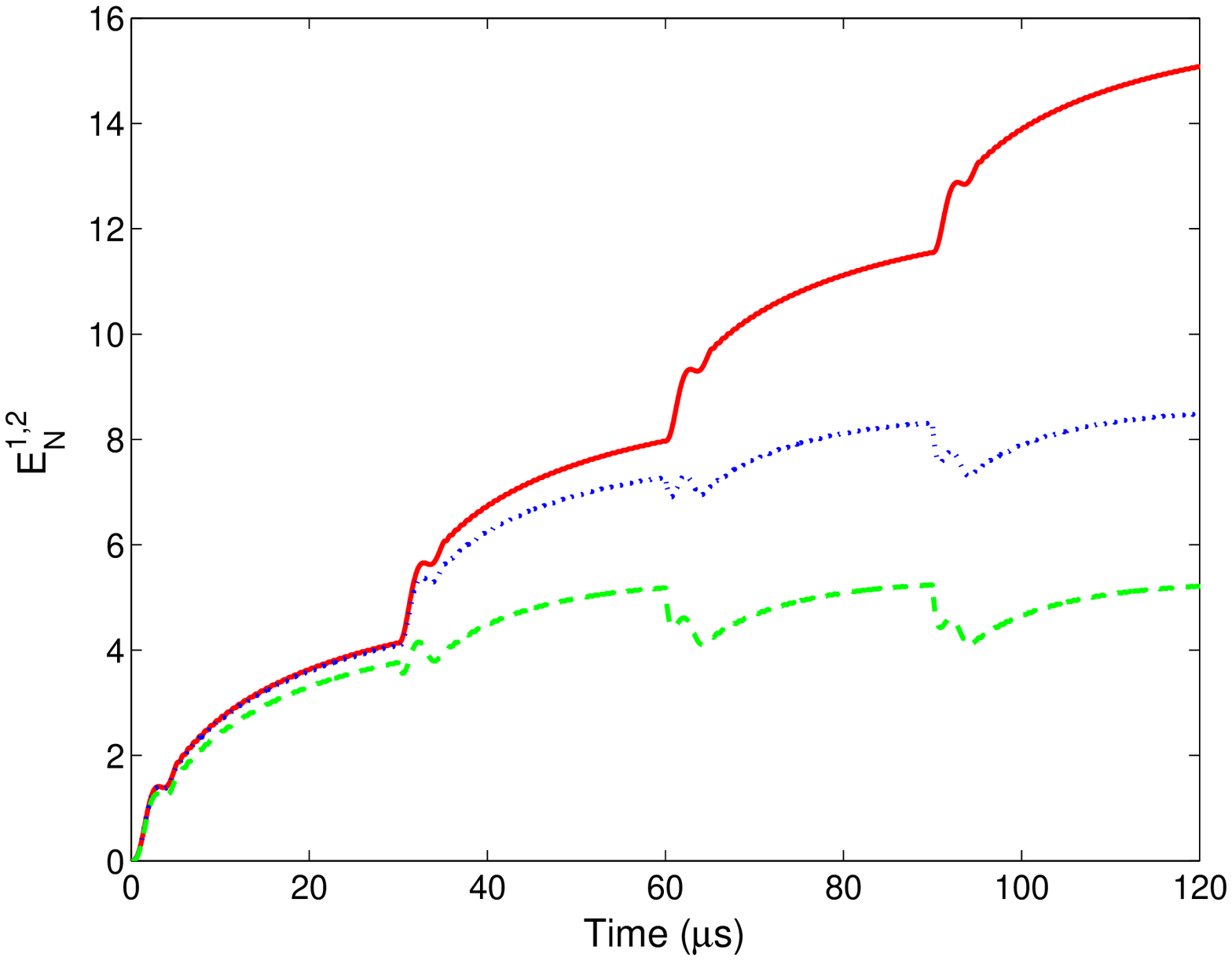}\label{ent_multistep}}
\caption{Dynamics of the entanglement (in ebits of logarithmic negativity)
between the radial modes of the ions, in a trap
containing $n=2$ ions and with longitudinal trapping frequency $\omega_{L}=1{\rm MHz}$.
In plot (a), the system starts off from the ground state for transverse trapping frequency
$\omega_{i}=20{\rm MHz}$ and then evolves under a constant trapping frequency $\omega_{f}=2{\rm MHz}$.
In plot (b), the initial frequency is $\omega_i = 20 {\rm MHz}$ and then
the system evolves for $5$ ${\mu}{\rm s}$ at frequency $\omega_{f}={2}{\rm MHz}$
and for $25$ $\mu{\rm s}$ at frequency $\omega_{i}$; for iterations of the cycle are displayed.
The continuous (red) curves refer to a case with no dissipation,
the dotted (blue) curves refer account for a heating rate $\epsilon=200{\rm Hz}$,
while the dashed (green) curves
refer to a case with heating rate $\epsilon=2{\rm KHz}$ (resulting from a coupling to the bath
$\gamma=10^{-4}{\rm Hz}$ and a temperature $T=294^{\circ}{\rm K}$).
\label{multienta2}}
\end{center}
\end{figure}

In the plots, we also report the entanglement of pairs of modes in the traps for four, three
and two ions.
For bipartite entanglement as well, as shown in Fig.~\ref{multienta2},
the degree of robust entanglement achievable is by far beyond the maximal values
obtained in quantum optical systems,\footnote[8]{To the best of our knowledge,
the highest {\em measured} value for the logarithmic negativity in optical systems
(inferred from state reconstruction) is $E_{\N}\simeq 1.6$ ${\rm ebits}$ \cite{french}.
A simple evaluation also shows that, exploiting the degree of squeezing
reported in
\cite{takeno07} and assuming {\em perfect}
beam-splitting operations, one could achieve at most $E_{\N} \lesssim 3$ ${\rm ebits}$.}
even considering rather modest frequency jumps.
The case portrayed in Fig.~\ref{ent_multistep}
starts from the ground state for $\tilde{\nu}_i=10{\rm MHz}$.
Afterwards, the frequency is switched
to ${\omega}_{f}=2{\rm MHz}$ for $5 \mu{\rm s}$ and then back to ${\omega}_i$
for $25 {\mu}{\rm s}$: such a cycle realizes a highly entangling operation
because it is tailored such that the squeezings act always in the same phase-space directions
and, in principle, can be iterated at will to obtain any desired degree
of entanglement. In practice decoherence and dissipation will degrade such an
entanglement after some iterations. However, as is apparent from
the plot, stable and robust entanglement (around
$10 \,{\rm ebits}$ of logarithmic negativity) between
two ions could be created under realistic conditions.

Also, in the traps with more than two ions,
the dynamics of the entanglement between pairs of modes displays a `monogamous'
behaviour \cite{monogamy}:
for four ions, when the entanglement between ions $1$ and $2$ fades, the entanglement
between $1$ and $3$ starts raising and then is replaced by entanglement between
$1$ and $4$, which in turn fades out before the revival of the quantum correlations between $1$ and $2$.
The entanglement between $2$ and $3$ -- strongly favoured by the fact that such ions are both
nearest neighbours {\em and} at perfect resonance with each other (not true in general, as
the Coulomb corrections on the trapping frequencies depend on the position in the trap) --
is always positive but, significantly, its peaks and dips follows those of the entanglement between
$1$ and $4$, as one would heuristically expect by virtue of monogamy.
As reminded above, the reported values of $\omega_{i}$ and $\omega_{f}$ represent the `bare'
trapping frequencies, not taking into account the corrections due to Coulomb repulsion: therefore,
the ions are not all at resonance during the evolution.
Now, if one assumes that local control were available, one could correct for such a mismatch
by applying different bare frequencies to ions in different positions along the trap's
longitudinal axis.
We applied such corrections in the four ions case considered above and found that, quite interestingly,
one can spread correlations more evenly in this way and
get closer to the ultimate bounds imposed by monogamy:
after $10 {\mu}{\rm s}$ (all the parameters being the same
aside from the corrections) one ends up with a state where each radial mode
is individually entangled with all the other radial modes (a situation which never arises without
corrections, see Fig.~\ref{multienta4}).

The continuous variable entanglement between two radial modes of the ions could be swapped to light
if cavities were added to the setup, with the potential to achieve unprecedented degrees of optical entanglement,
as the parametric processes are definitely outperformed by the strategy outlined above based on the
trapping frequency's control: this possibility will be the subject of
a detailed investigation in a forthcoming paper \cite{noialtri}.

\section{Propagation of quantum information}\label{propa}

The locally defined radial modes of the ions, interacting via the
Hamiltonian (\ref{hamil}), are an example of {\em harmonic chain},
where the coupling between different oscillators can be, to a very
good extent, controlled. The propagation of quantum information in
such a setting has been proposed and analysed theoretically in
\cite{martinhartley}, but not yet realised in practice since mechanical oscillators 
have yet to enter the quantum regime 
(due to the technical difficulties still encountered in controlling
both the couplings and the decoherence and dissipation of
chains of nano-oscillators \cite{nanoexp}). Trapped ions could thus provide
a very promising alternative to implement harmonic chains, realize
quantum data buses \cite{martinsemiao} and test
the results predicted by theoretical studies. More generally, such
a demonstration would be a further application of ion traps as
{\em quantum simulators}, as already proposed in several past
studies \cite{porras2004ab,retzkerphtr}. In this Section, we shall
present an overview of the possibilities offered by the radial modes
as harmonic chains for the propagation of quantum states, with
quantitative investigations, in scenarios similar to those detailed
in the preceding sections.

\subsection{Transmission of two-dimensional quantum states}\label{singlex}

\begin{figure}[t!]
\begin{center}
\subfigure[a]
{\begin{tabular}{cc}\label{Prop10_50_50}
\includegraphics[scale=0.18]{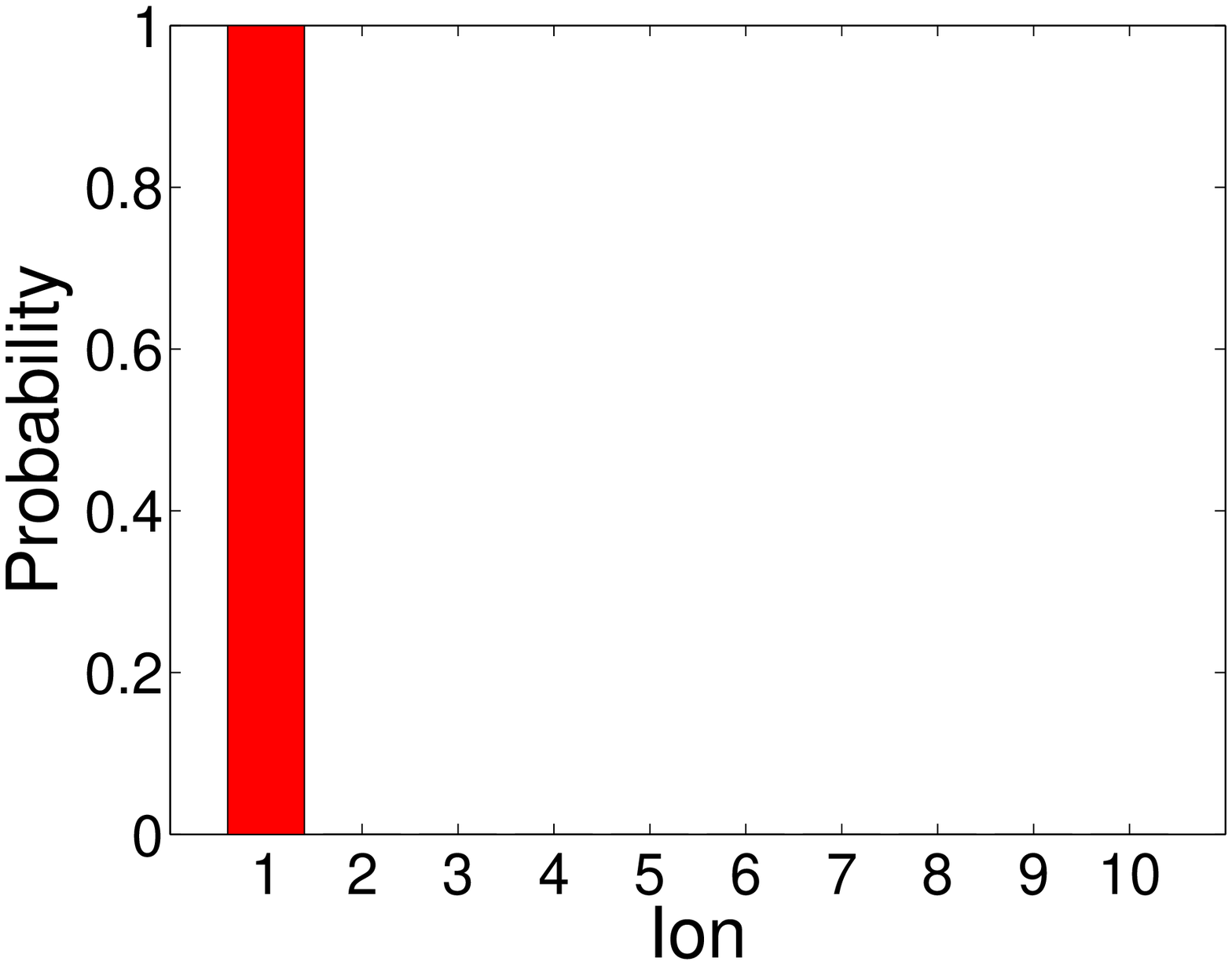}\\
\includegraphics[scale=0.18]{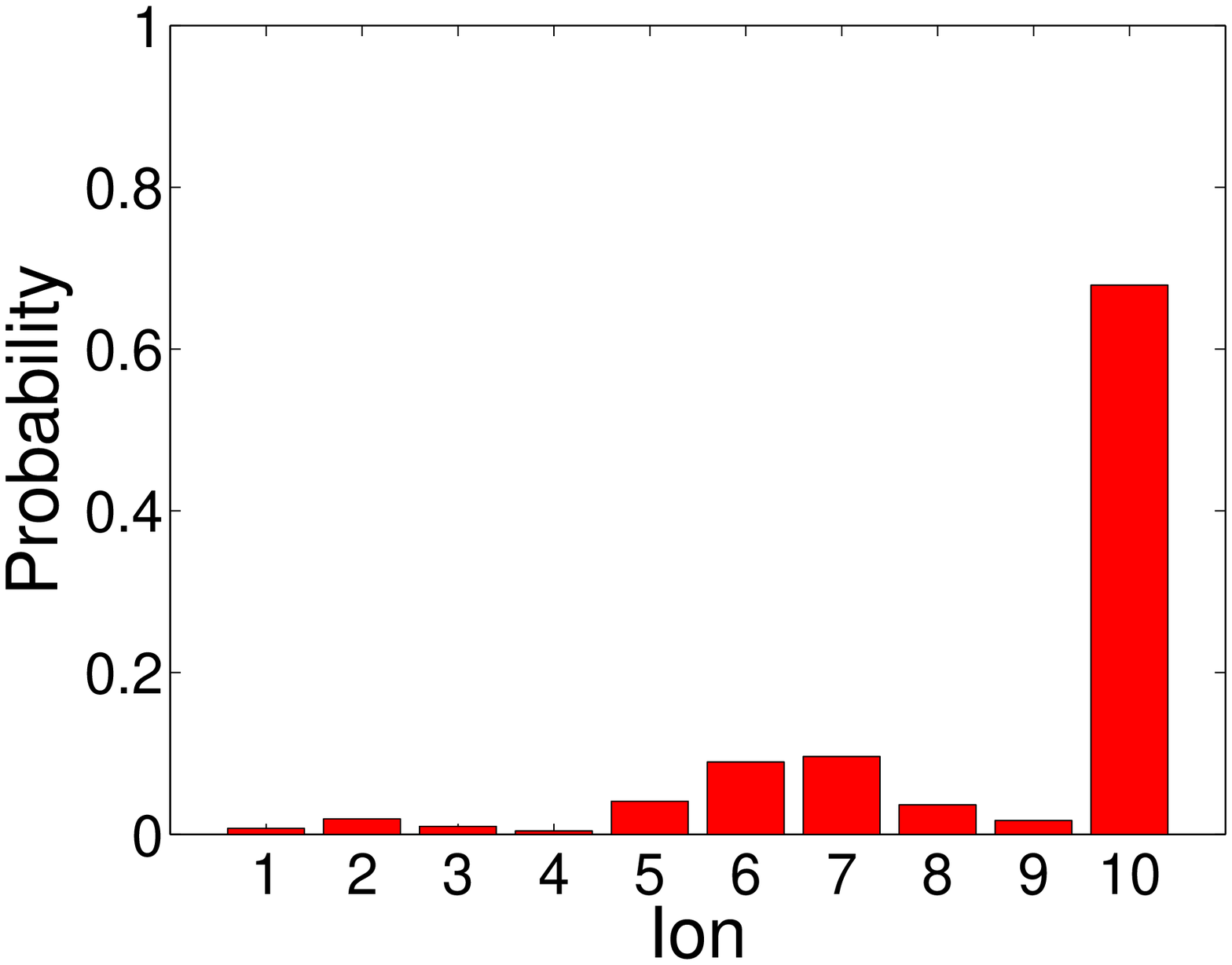}\\
\includegraphics[scale=0.18]{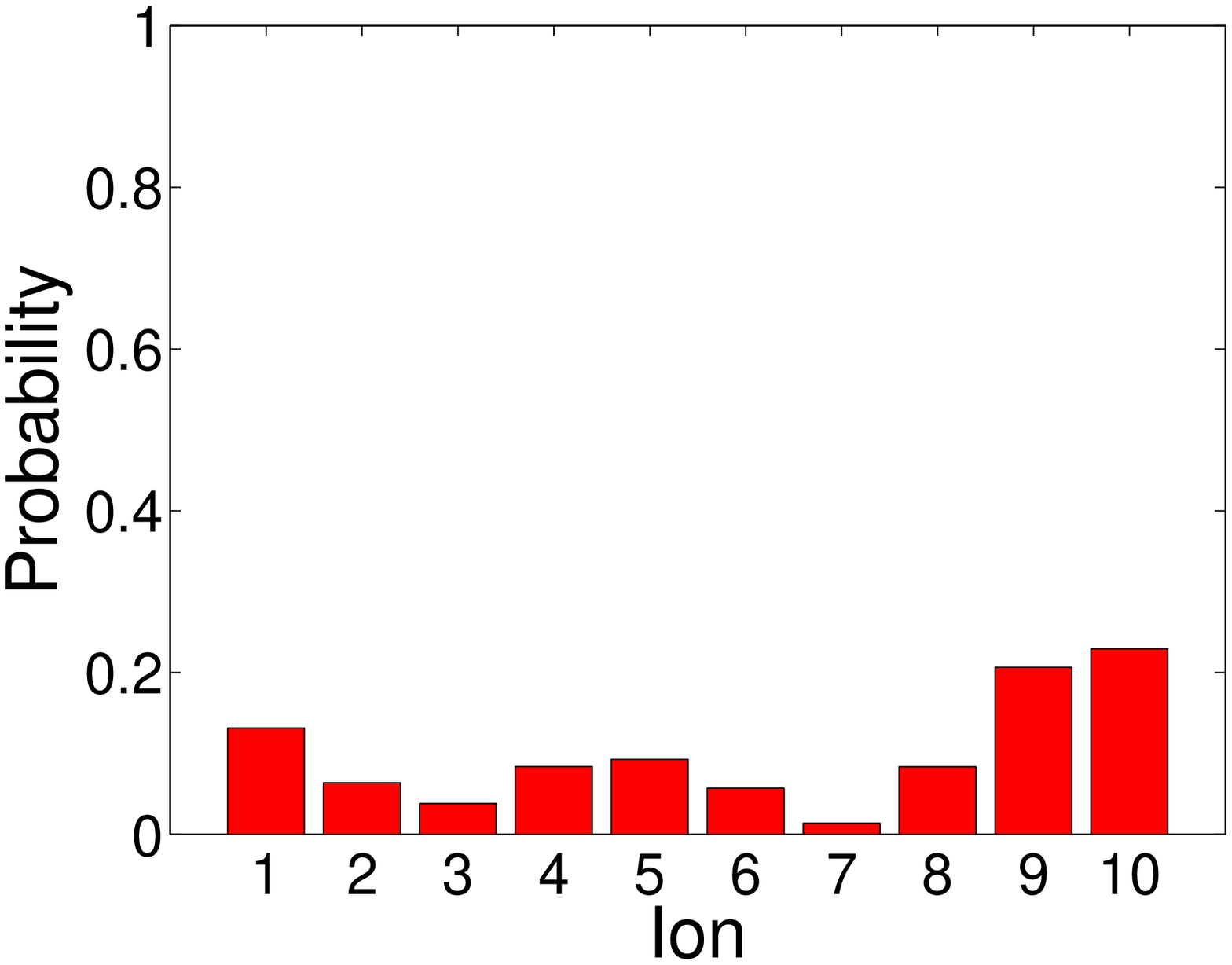}
\end{tabular}}
\subfigure[b]
{\begin{tabular}{cc}\label{Prop10_50_5}
\includegraphics[scale=0.18]{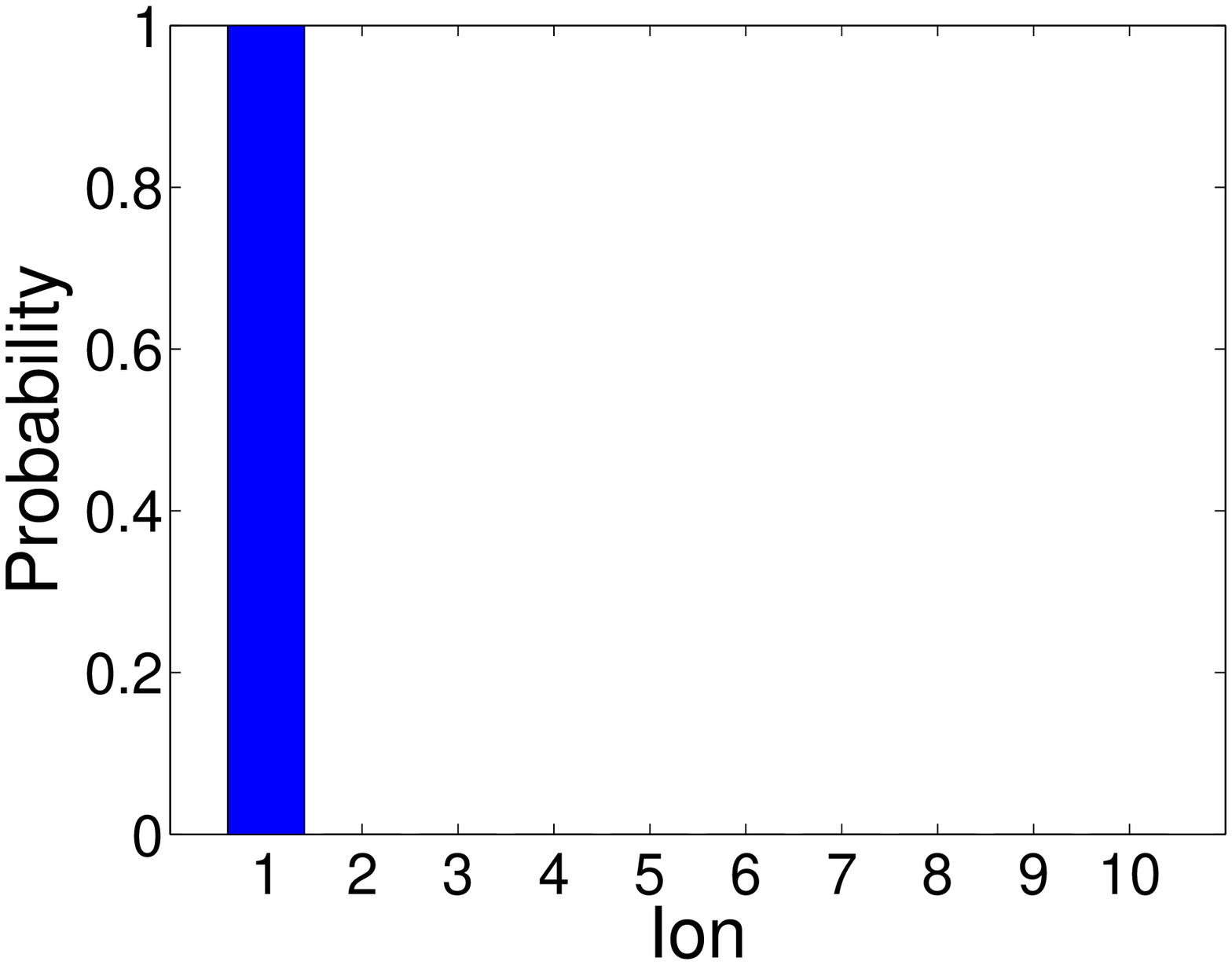}\\
\includegraphics[scale=0.18]{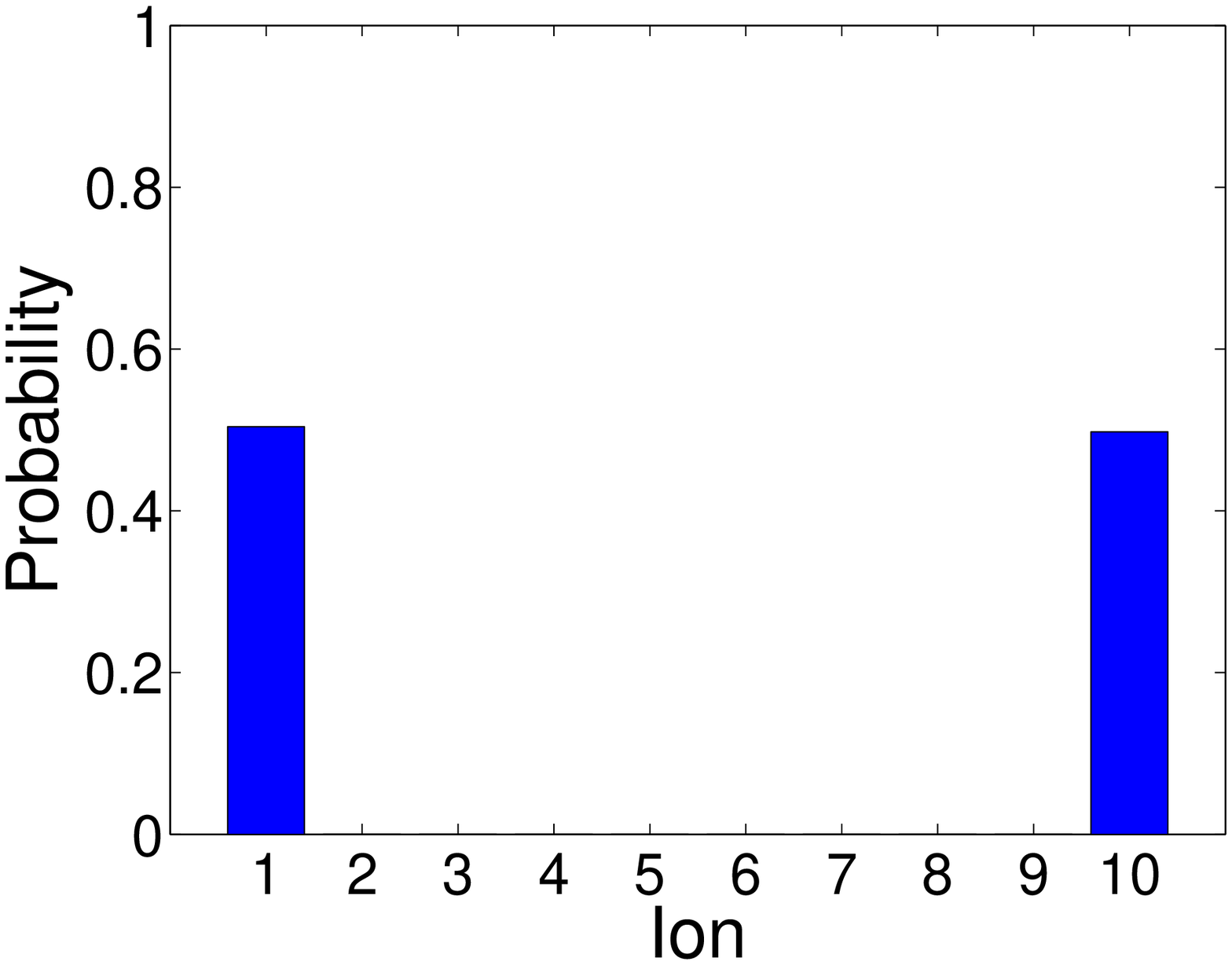}\\
\includegraphics[scale=0.18]{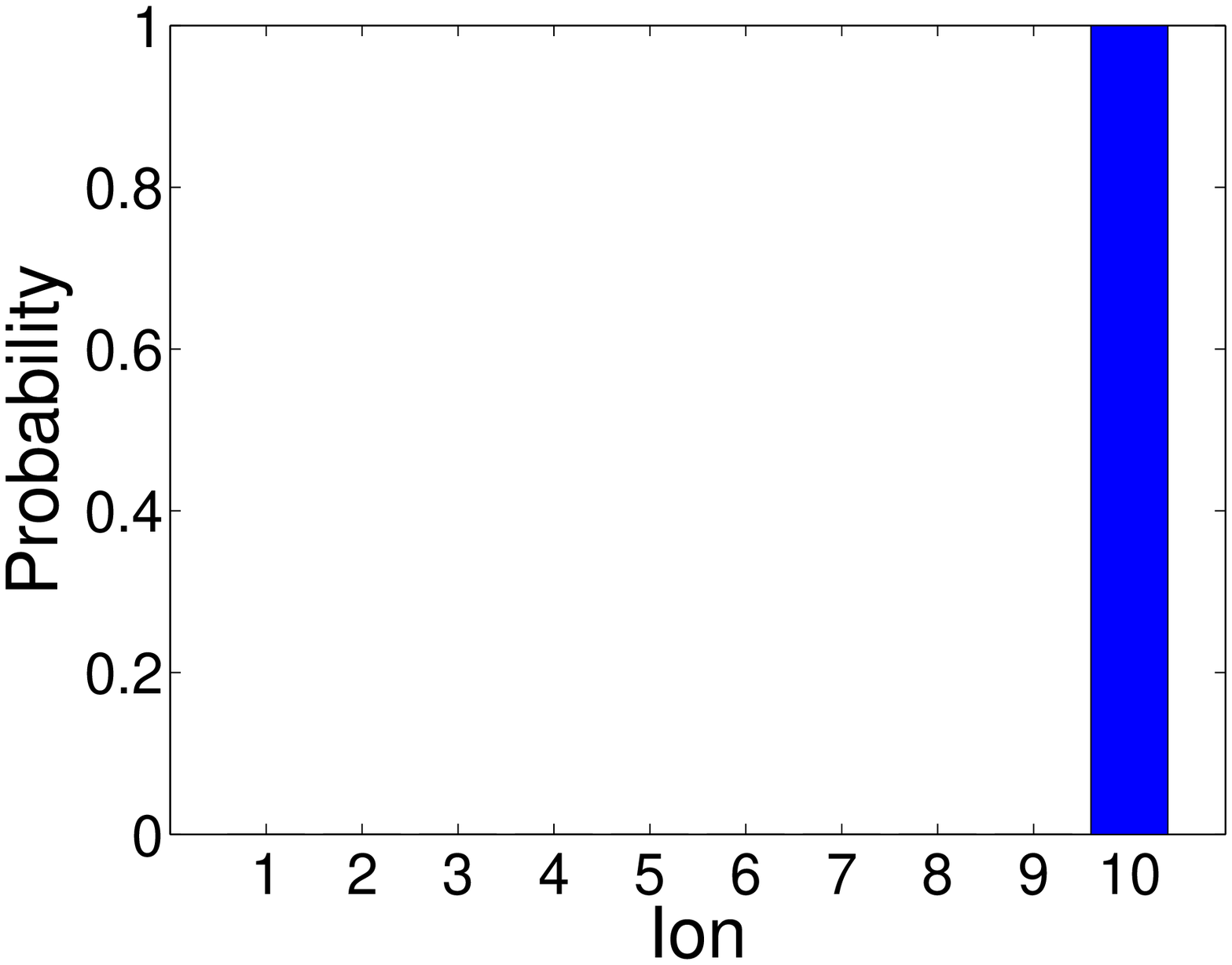}
\end{tabular}}
\subfigure[c]
{\begin{tabular}{cc}\label{Prop10_50_10glob}
\includegraphics[scale=0.18]{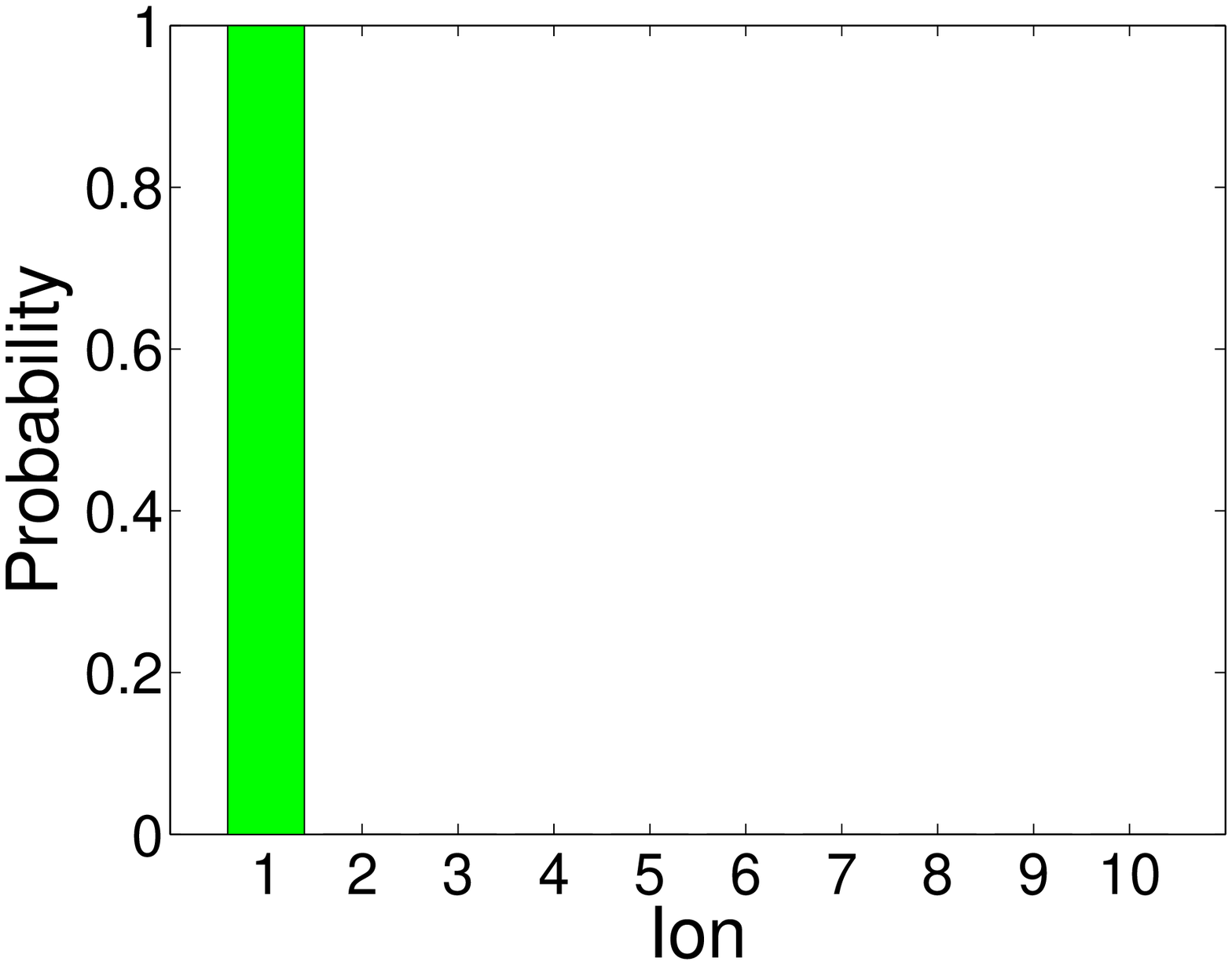}\\
\includegraphics[scale=0.18]{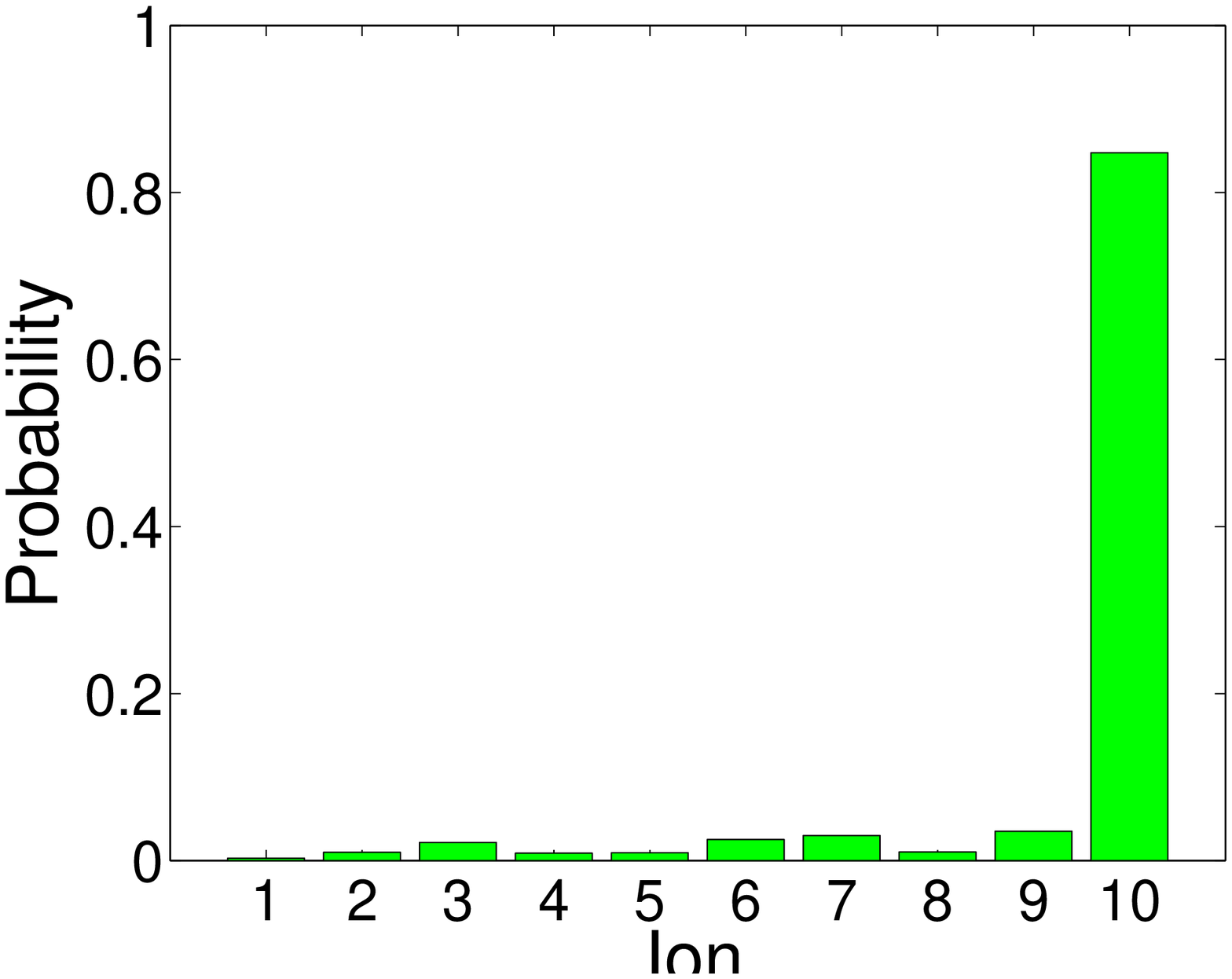}\\
\includegraphics[scale=0.18]{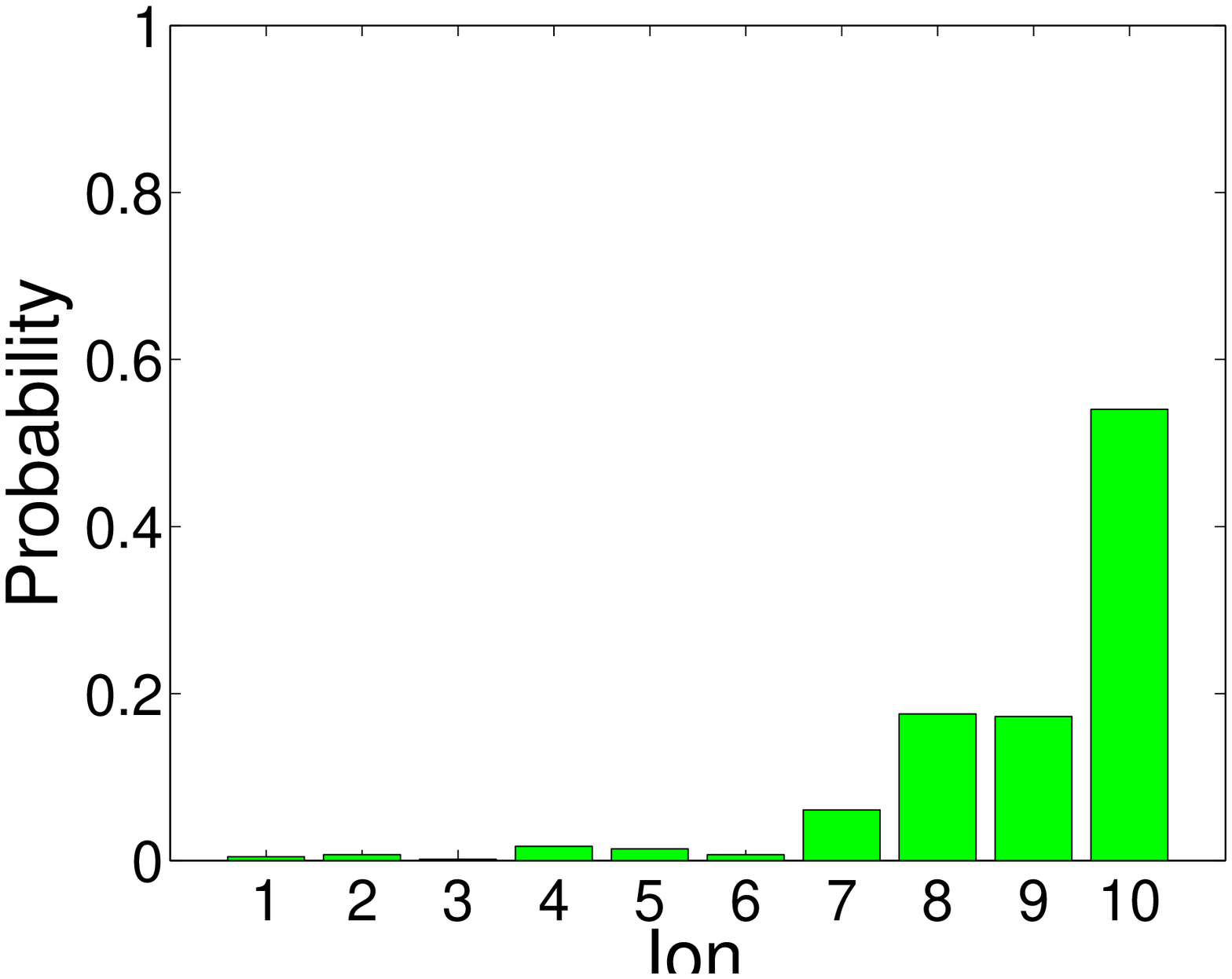}
\end{tabular}}
\caption{Leftmost histograms: probability amplitude of a single excitation (phonon) across the chain,
if the evolution occur for frequencies $\omega_{j}=50\omega_{L}$ $\forall j$ after times (from top to bottom)
$t=0$, $t=375\omega_{L}^{-1}$ and $t=5600\omega_{L}^{-1}$.
Middle histograms: probability amplitude of a single excitation (phonon) across the chain,
if the evolution occur for frequencies $\omega_{1}=\omega_{10}=5\omega_{L}$
and $\omega_{j}=50\omega_{L}^{-1}$ for $j\neq1,10$ after times (from top to bottom)
$t=0$, $t=2800\omega_{L}^{-1}$ and $t=5600\omega_{L}^{-1}$.
Rightmost histograms (red): probability amplitude of a single excitation (phonon) across the chain,
if the evolution occur for frequencies $\omega_{j}=10\omega_{L}$ $\forall j$ after times (from top to bottom)
$t=0$, $t=625\omega_{L}^{-1}$ and $t=640\omega_{L}^{-1}$.}
\end{center}
\end{figure}

The transmission of quantum information encoded in
finite dimensional quantum states (most often qubit states) through a chain of
interacting quantum systems has drawn much attention in recent years \cite{transfer},
as such systems are envisaged as possible quantum buses, linking
different parts of future quantum processors. However, despite
considerable efforts, finding systems where such a transmission can
be realised with sufficiently small level of noise has proven to be
very challenging. Therefore, good practical candidates for such tasks
are still of interest. Proposals in this context have so far
focused on the transmission of finite dimensional states in chains of
interacting (effective or proper) `spins'. Here, instead, we consider qubit
states encoded in the bosonic Hilbert space of the radial modes of
trapped ions and, guided by the results of the Section \ref{manip},
we show that such modes are able to send qubit states through chains
of trapped ions.

As a preliminary remark, let us notice that, if the ratio between the radial and the longitudinal
trapping frequencies is very large, $\omega_{j}/\omega_{L}\gg1$ $\forall j$, then
the effect of the Coulomb interaction on the Hamiltonian governing the radial modes
is negligible\footnote[9]{See page \pageref{uus} and notice that, quite remarkably, the ratios
$d/|u_j-u_k|$ does not depend on the trapping frequencies nor on the masses of the ions
but only on the total number of ions $n$ (see \cite{james98} for details).}
and the Hamiltonian (\ref{hamil}) reduces to $\hat{H} \simeq \frac12 \hat{R}^{\sf T} \gr{\tilde{\omega}}
\oplus \gr{\tilde{\omega}} \hat{R}$.
That is to say, the radial oscillations of the
individual ions are decoupled, with local normal frequencies approximately given by $\{\omega_{j}\}$.
In this situation, the ground state of the system is just the ordinary, non-squeezed, vacuum: a Gaussian
state with CM equal to the identity and vanishing first moments. For instance,
for $\omega_j=50\omega_{L}$ $\forall j$ the relative discrepancy between the actual ground state
and the vacuum is only $10^{-3}$ in terms of covariances.
Starting from such a ground state,
we will consider a scenario where the sender has its qubit state encoded in the single excitation sector
of the oscillations of the first ion ({\em i.e.}, on the subspace spanned by the phonon states
$\ket{0}$ and $\ket{1}$) and wants to send the excitation through the chain to a receiver who
owns the last ion.
Notice that, as recalled in Sec.~\ref{further},
a (non-Gaussian) single qubit state of this kind could be prepared, in practice,
by coupling the motion of an ion to the internal degrees of freedom
through a proper sequence of blue and red sideband pulses.

Neglecting, for the time being, thermal noise (see below), one has that the state $\ket{0}$
can be sent perfectly through the chain. Therefore, the overall fidelity of the transmission
depends solely on the fidelity with which $\ket{1}$ can be sent from the first to the last ion.
The amplitude of such a transfer can be determined analytically by tracking the evolution of the
initial operator $a_{1}^{\dag}$ (creating the excitation in the first ion) in Heisenberg picture.
Because the system is harmonic the evolution of $a_{1}^{\dag}$ results in a linear combination
of field operators, which can be determined with standard techniques. We could thus consider
different situations.
The histograms we plot show the probability amplitude $P$ for finding a single excitation at
different sites in the chain. The resulting transmission fidelity
(averaged over the Haar measure of a single qubit space) is given by
$1-\frac12 (1-\sqrt{P})^2+\frac34 \sqrt{P}(1-\sqrt{P})$, monotonically increasing with $P$.
We will consider a chain of $n=10$ ions and an initial ground state for trapping frequencies
$\omega_{j} = 50\omega_{L}$ $\forall j$.

Fig.~\ref{Prop10_50_50}. shows the case where the trapping frequencies
are left unchanged after the initial preparation.
Because the interacting terms, while very small, are still present, the excitation propagates because
the local mode of the first ion is not exactly a normal mode of the system.
After an evolution time $t=375\omega_{L}^{-1}$, $0.76$ of the probability amplitude is transferred
to the final state.
However, the propagation is rather dispersive, as each mode
is on resonance with all the others.

In order to boost the probability of transmitting the excitation, sender and receiver can switch the
local trapping frequencies of their respective ions from
the initial value $\omega_{i}=50 \omega_{L}$ down to a common value $\omega_{f}$:
in this way they can realise the beam splitting operation at distance described in Sec.~\ref{bs}
which, in principle, would allow for perfect swapping. This specific example thus also
illustrates our previous general discussion and shows with which precision and over which operating times
are linear operations actually possible over a chain of ions.
As one can see from Fig.~\ref{Prop10_50_5}, where $\omega_{f}=5\omega_{L}$ has been assumed,
the beam splitting operation is virtually perfect: all the
probability amplitude is gradually transferred to the final ion, while the other ions are never involved in the
process. This quality in the transfer comes at the expense of transfer time: over $10$ ions
the beam splitting operation takes roughly an order of magnitude more ($t\simeq5600\omega_{L}^{-1}$)
than the imperfect transfer considered
in Fig.~\ref{Prop10_50_50}. This is simply due to the fact that, by keeping the middle ions on resonance,
one takes advantage of the interaction between nearby ions, which are clearly stronger
(as the interaction decays like the cube of the distance) and propagate the excitation through the chain.
For given trapping frequencies, the time needed to achieve the beam splitter between
radial modes at the two end of a chain of $n$ ions scales very accurately as $n^2$.
Notice also, as a side remark, that since the Coulomb corrections to the local trapping frequencies are
symmetric with respect to the longitudinal centre of the trap, the transfer between extremal ions is
somewhat favoured in practice, as equal trapping potentials will result in equal effective trapping frequencies
for such ions.

Finally, in Fig.~\ref{Prop10_50_10glob},
we consider a `compromise' between the two cases analysed above:
all the frequencies are changed to the frequency $10\omega_{L}$
obtaining, after a time $t=625 \omega_{L}^{-1}$,
a transferred probability amplitude of $0.85$.
It is important to remark that the slower beam splitting operation, where
the quantum information does not disperse through the chain at all, is not only more precise
than the other two options examined,
but also much
more stable with respect to imperfections in the allowed interaction times.

Let us now turn to inspect the effect of decoherence:
clearly, thermal phonons are deleterious for the transfer of information encoded in
single excitation sectors. The analytical estimates for the transfer times we just presented allow us to
determine the restrictions on the heating rates that would allow such systems to transfer information
in practical situations. Assuming a longitudinal trapping frequency of $1$ ${\rm MHz}$ and a
(less ambitious) chain of $5$ ions, one would realise a beam splitting operation (perfect transfer)
between extremal ions in $t\simeq1.4 {\rm msec}$. In such a case, heating rates as low as
$\epsilon\simeq0.1$ ${\rm kHz}$ would be needed for a coherent transfer to take place.
For such heating rates, the action of the thermal phonons on the fidelity between initial and final state
can be assumed to be linear resulting, after average over the Haar measure,
into a mean fidelity ${\cal F}\simeq 1- 3\epsilon t/2\simeq 0.8$.
\footnote[0]{The effect of losses and thermal phonons
for small $\epsilon$ (so that $\epsilon t \ll 1$ over the interesting time-scales)
can be easily estimated by letting the master equation \eq{master} for $N\gg1$ act
at first order on an otherwise perfectly transferred generic
state $\alpha\ket{0}+\beta\ket{1}$, and by setting $\epsilon=\gamma N$.
The fidelity between initial and final states (corresponding to the overlap, for pure initial state) can be then
determined and averaged over the Haar measure
of a single qubit Hilbert space, to obtain the mean value ${\cal F}\simeq 1 - 3 \epsilon t/2$.}
Such heating rates are certainly demanding for arrays of traps, but not inconceivable,
considering the heating rates achieved for single traps and the fact that current systems all operate
at room temperature (thus, in principle, two orders of magnitude could be gained on heating rates).

\subsection{The role of the internal degrees of freedom}

In the preceding section, while discussing the transfer of finite dimensional
quantum information,
we started off from a situation where a qubit state is encoded in the first
two number states of the first ion.
As already mentioned above, one of the most expedient ways to create such a
finite dimensional state for the motional degrees of freedom
is creating the desired state for
the `internal' degrees of freedom of the ion
(embodied by its internal electronic levels)
and then coupling them to the local radial modes to achieve the swapping.
In order to provide the reader with a more comprehensive treatment and
give him/her a flavour of the way the internal degrees of freedom
enter in such dynamics, we briefly give account of a specific example
where the internal degrees of freedom are involved
throughout the whole transmission
of quantum information.

\begin{figure}[t!!]
\begin{center}
\includegraphics[scale=0.37]{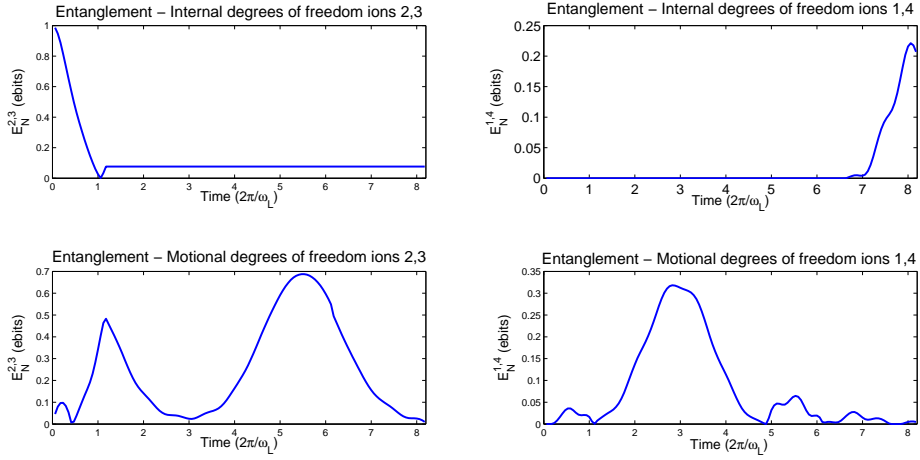}
\end{center}
\vspace*{-0.5cm} \caption{Propagation of the entanglement (logarithmic negativity)
between internal and motional degrees of freedom.
The initial
state is a Bell singlet between the internal degrees of freedom of the second and the third ions;
the transverse trapping frequency is $10\omega_{L}$.
The upper (lower) plot on the left column shows the logarithmic negativity between
the internal (motional) degrees of freedom of ions $2$ and $3$.
The upper (lower) plot on the right column shows the logarithmic negativity between
the internal (motional) degrees of freedom of ions $1$ and $4$.
The dynamics is divided to three stages.
In the first stage (for $0\le t\le1.2\frac{2\pi}{\omega_L}$),
the coupling between internal and motional degrees of freedom of ions $2$ and $3$ is on, and
the entanglement is swapped the former to the latter.
In the next stage (for $1.2\frac{2\pi}{\omega_L}\le t\le6.2\frac{2\pi}{\omega_L}$),
all the internal and motional degrees of freedom couplings are off, and the entanglement propagates
through the chain via the motional degrees of freedom.
Finally, for $t\ge6.2\frac{2\pi}{\omega_x}$, the
the coupling between internal and motional degrees of freedom of ions $1$ and $4$ is on, and
the final swapping takes place.}\label{six1}
\end{figure}

We shall consider a trap with four ions, with $\omega_{j}=10\omega_{L}$ for $1\le j\le 4$.
To keep the example down to earth, we will not
assume any capability of local control nor any change during the evolution
for the radial frequencies here.
The only manipulation required will be the switching on and off of the
coupling between the local phonons and the two levels of each ion
where the initial qubit state is encoded.
Denoting such levels $\ket{e}_j$ and $\ket{g}_j$, and
setting $\sigma_{x,j} := \ket{e}\bra{g}_{j} + \ket{g}\bra{e}_{j}$, we shall
assume the following interaction Hamiltonian $\hat{H}^{int}_j$ for each ion $j$:
$$
\hat{H}^{int}_j = g_{j} \sigma_{x,j} \hat{X}_j \; ,
$$
which describes appropriately the coupling realised in experiments
(see, {\em e.g.}, \cite{Meekhof1996}).
For simplicity, the interaction strengths $g_j$
will be all set to $\omega_L$ (notice that, because the oscillators' frequency are set to $10\omega_L$,
the interaction reduces, up to a very good approximation, to a rotating wave one).
As initial state, at $t=0$, we assume a Bell state between the
internal degrees of freedom of ions $2$ and $3$, and the ground state for the remainder f the system:
$$\ket{g}_1\otimes(\ket{g}_{2}\otimes\ket{e}_3+\ket{e}_2\otimes\ket{g}_3)\otimes\ket{g}_4 \otimes \varrho_g \, ,$$
where $\varrho_g$ is the ground state of the radial modes of the ions.
As a signature for the transmission of quantum information,
we will consider the evolution of the entanglement
between the internal and motional degrees of freedom
of ions $2-3$ (initially entangled) and ions $1-4$
(which gets entangled through the process), which is reported in Fig.~\ref{six1}
in tems of logarithmic negativity (the logarithmic negativity between internal degrees of freedom
of ions $j$ and $k$ is denoted $\tilde{E}_{\N}^{j,k}$).
The dynamics is then split into three stage:
from time $t=0$ to $t=1.2\cdot2\pi/\omega_L$ the
internal and motional degrees of freedom of ions $2$ and $3$
are coupled and the entanglement is dynamically swapped from
the former to the latter. Next, from $t=1.2\cdot2\pi/\omega_L$
to $t=6.2\cdot 2\pi/\omega_L$, the coupling is switched off and
the entanglement propagates through the chain via the motional degrees of freedom,
and the radial modes of ions $1$ and $4$ gets entangled.
Finally, from $t=6.2\cdot 2\pi/\omega_L$ to $t=10\cdot 2\pi/\omega_L$,
the entanglement is swapped to the internal degrees
of freedom of ions $1$ and $4$.
One can see that, though the quantum information partially disperses under such
conditions (essentially due to the stringent restrictions we put, in this instance,
on the control of the local frequencies),
the process is capable of transferring the entanglement from the internal
degrees of freedom of ions $2$ and $3$ to those of ions $1$ and $4$
through the radial motional degrees of freedom.


\subsection{Propagation of continuous variable states and entanglement}

Clearly, quantum information (and entanglement) can be propagated
through the chains under examination also at a continuous variable
level, when populating the whole infinite dimensional Hilbert space.
The study of information and entanglement propagation over harmonic
chains has been proposed and discussed in detail in
\cite{martinhartley,martinsemiao}.
Here, we shall investigate cases of such propagations along the chains of ions we are considering,
addressing specific instances which can be realised in the laboratory.
Our aim is pointing out at what degree and under what conditions can the theoretical schemes
based on harmonic chains be implemented on radial modes of trapped ions.

As argued in \cite{martinhartley}, the capacity of transmitting
quantum states between distant parties is closely related to the
capacity of transmitting, or ``swapping'', entanglement between
them. In fact, the latter also critically requires the transfer
to happen coherently through the whole process. In view of this
fact we will limit ourselves to consider the propagation of CV
entanglement (rather than the fidelity between sent and received
quantum states). Also, in order to focus on a feasible scenario,
we shall consider a common initial condition: the chain of $n$
ions starts off from the (completely separable) ground state for
$\omega_{j}=\omega_{0}$ $\forall j$ and, then, entanglement between
the first and the second ion is created, as described in
Sec.~\ref{enta}, by switching their frequencies to the same
frequency $\omega_{e}$. From now on the first ion is left off-resonance
at frequency $\omega_{e}$ (and thus effectively isolated from
the chain), and the aim is transmitting its initial entanglement
with the second ion through the chain over to the final $n-$th
ion. Two options will be examined:
\begin{itemize}
\item[i)]{the entanglement is swapped by a chain of beam splitters between neighbouring ions
(achieved by setting such neighbouring ions at the original frequency while putting the others
off-resonance);}
\item[ii)]{the entanglement is swapped directly by a beam splitter between the ions at the far ends of the chain.}
\end{itemize}
Notice that in case i) the beam-splitters operate
at the original frequency $\omega_{0}$ so that no local squeezing takes place,
as evident from \eq{simple}.
Local squeezing would affect the entanglement between the modes but also, crucially, alter the original state
to be transmitted and is thus undesirable in the present context.

\begin{figure}[t!]
\begin{center}
\subfigure[a]
{\includegraphics[width=6.3cm,height=7cm]{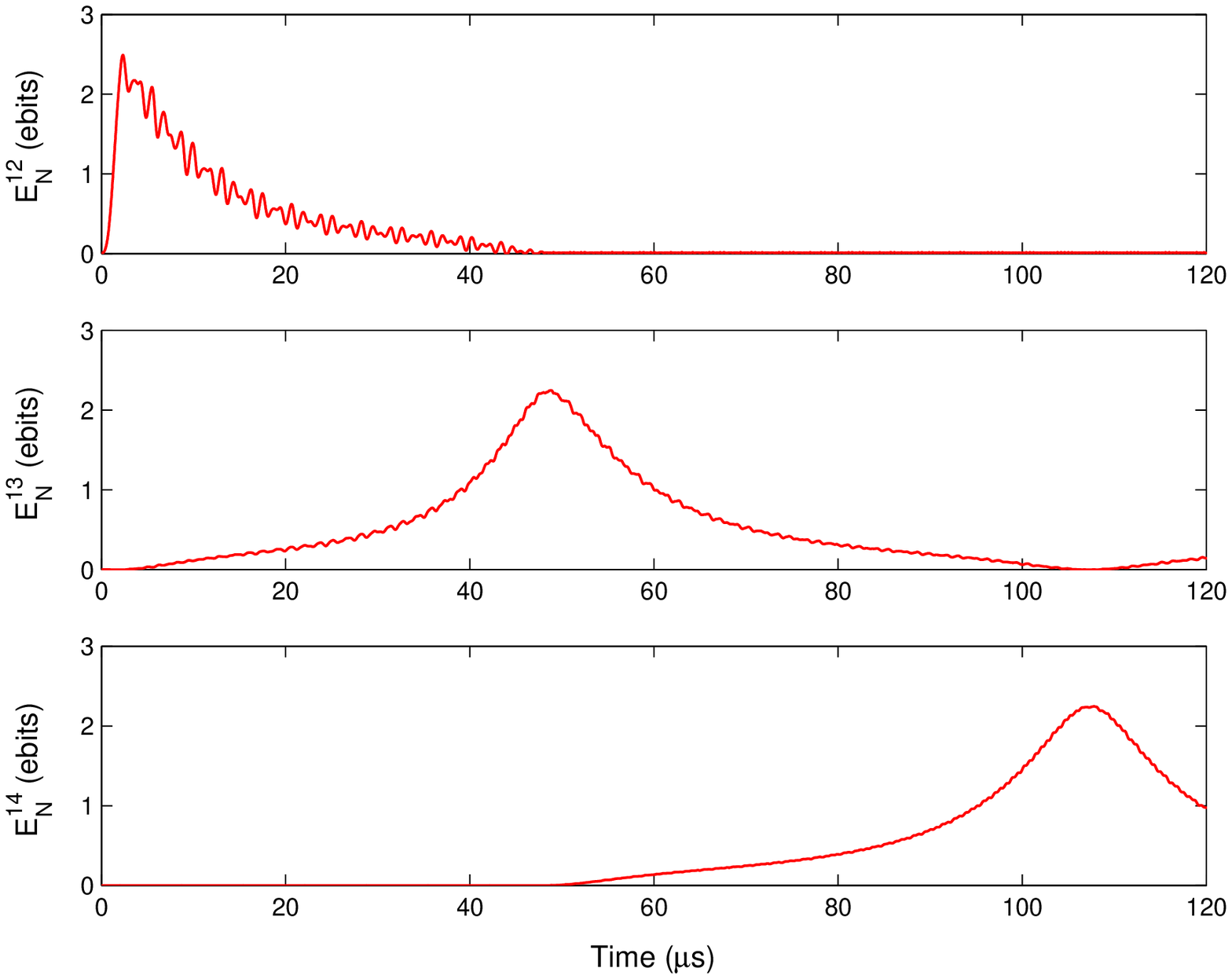}}
\subfigure[b]
{\includegraphics[width=6.3cm,height=7cm]{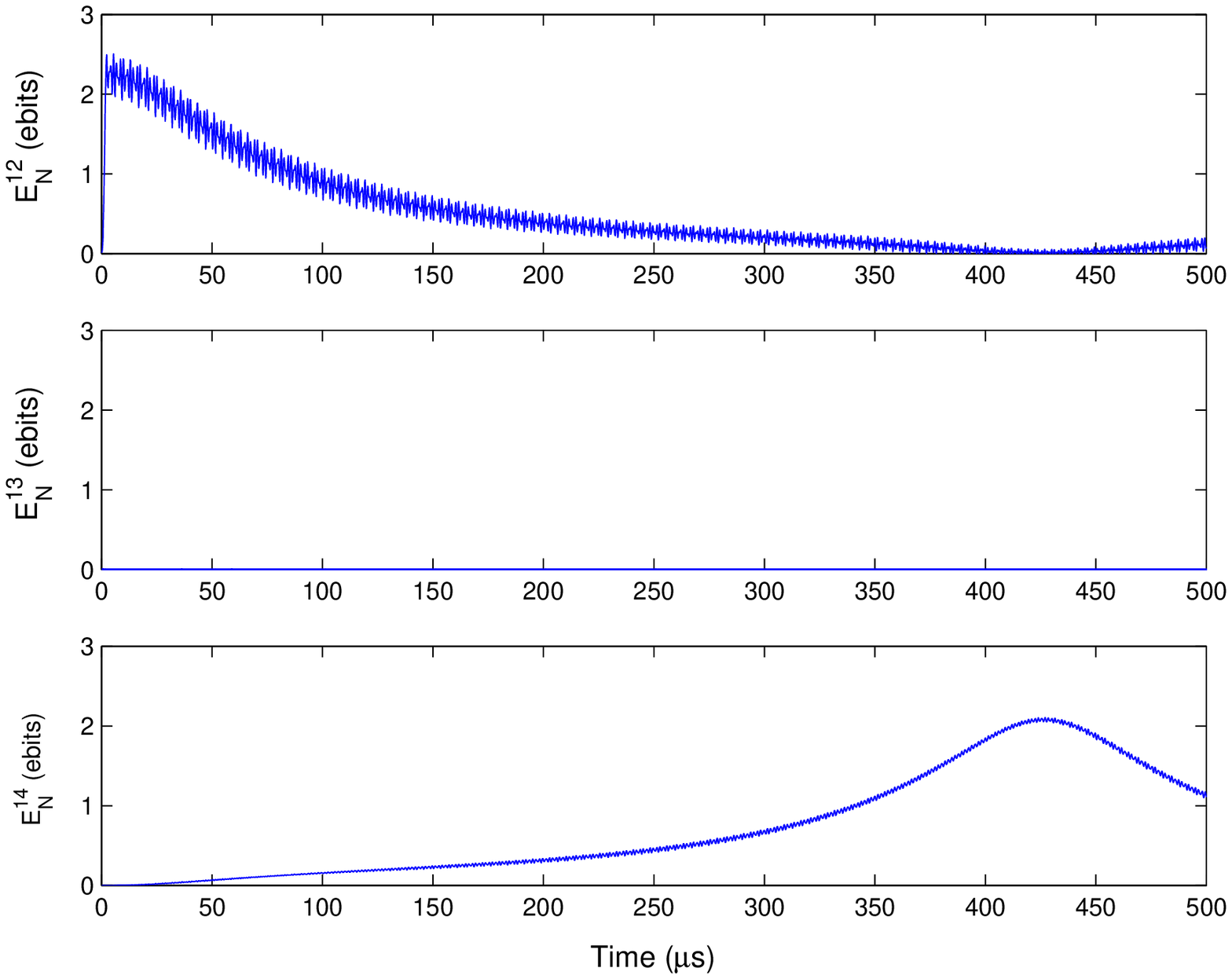}}
\caption{Transfer of the entanglement with the first ion of a chain (measured by ebits of logarithmic negativity
$E_{\N}^{jk}$)
from the second ion to the last one, in a chain of four ions. From top to bottom, the plots
show the entanglement between ions $1$ and $2$ ($E_{\N}^{12}$), $1$ and $3$ ($E_{\N}^{13}$)
and $1$ and $4$ ($E_{\N}^{14}$). On the left side (a), the entanglement is transferred by a relay of
beam splitters obtained by switching the radial trapping frequencies of the ions $2$ and $3$
to $\sqrt{98.8}\omega_{L}$
\label{entprop} for a time $46\omega_{L}^{-1}$ while keeping the other ions off-resonant and then, finally, by
switching the trapping frequencies of ions $3$ and $4$ to $\sqrt{98.8}\omega_{L}$
(with the other ions off-resonant).
On the right side (b), the entanglement is transferred directly by a beam-splitter between ions $2$ and $4$,
obtained by setting their frequencies to $\sqrt{98.8}\omega_{L}$ while keeping the others off resonance.
In both cases, the state at time $t=0$ is the ground state for all trapping frequencies equal to $\sqrt{98.8}\omega_{L}$
(such a value takes into account the corrections due to Coulomb repulsions) and,
during the initial $2$ $\mu{\rm s}$, the entanglement between ions $1$ and $2$ is built up
by bringing their `bare' trapping frequencies down to $2$ $\mu{\rm s}$.}
\end{center}
\end{figure}

The two strategies i) and ii) are compared in Fig.~(\ref{entprop}) for a chain of four ions.
Such a small system already permits one to highlight all the essential features
of the two strategies.
As apparent, a complete dynamical swapping of the entanglement can be achieved in both cases,
with remarkable accuracy.
A chain of beam splitters allows for a faster transfer, as one should expect since the coupling $\kappa_{jk}$
between distant ions is inversely proportional to their cubed distance: in general,
the time needed to send information across $n$ ions by a `relay' of beam splitters roughly scales like
$n$, whereas the time needed to achieve a beam-splitting operation between first and last ion scales like $n^3$.
A `relay', taking less time, is thus less sensitive to decoherence and dissipation.
On the other hand,
a single beam-splitting operation between distant ions only requires a single adjustment of the trapping
frequencies (after the initial entanglement is created),
whereas a chain of beam-splitters requires $n-2$ manipulations, which would be possibly difficult to master
in practice and could involve more errors and imperfections.
Therefore, we have investigated the effect of imperfections in frequencies and evolution times
on the transfer for both cases: the series of beam-splitters turns out to be more robust.
In the instance depicted in Fig.~\ref{entprop},
the transfer by a series of beam-splitters is left virtually unaffected
by uncertainties of $0.1\omega_{L}$ on frequencies and $0.1\omega_{L}^{-1}$ on operating times
($\omega_L$ is the longitudinal trapping frequency) whereas, on average,
the entanglement transferred by a single beam splitter is reduced to
to $\simeq 1.2$ ebits of logarithmic negativity (out of $\simeq 2.2$ ebits initially present between ions $1$ and $2$)
by the same imperfections.
Summing up, strategy i) proved to be more reliable and faster (and thus less subject to decoherence
and dissipation)
and is hence to be preferred for the transmission of CV quantum information over chains of ions.
Considering once again the example of Fig.~\ref{entprop}, for a longitudinal frequency $\omega_{L}=1$ ${\rm MHz}$,
an effective coherent transfer would take about $100$ $\mu{\rm s}$ and thus require heating rates
around $1$ ${\rm KHz}$ to be carried out effectively.

\section{Nonlocality tests} \label{bell}

The presence of strong Gaussian multipartite entanglement highlighted
in Sec.~\ref{enta} can be experimentally demonstrated and put to use
in testing quantum nonlocality.
Central to this endeavour is
the capability of performing non-Gaussian measurements on the
motional state of the ions, pointed out in Sec.~\ref{further}.
In particular, we already recalled that phonon-numer parity measures
are possible on single copies of the system, and so are displacement operations.
In this section, we will explore the possibility of violating multipartite Bell-like
inequalities (the so-called ``Klyshko'' inequalities \cite{klyshko}) by
measuring ``displaced parity'' observables, as proposed for generic CV systems
by Banaszek and Wodkiewicz \cite{banaszek98}.
Although bound to be subject to the locality loophole
(considering that the distances between the ions are typically of the order of $1 {\rm \mu}m$),
such an experiment would be a remarkable test of quantum non-locality with massive
particles, which is still lacking.

To fix ideas and address a situation within the reach of current experiments,
we shall study the test on the three-mode Gaussian state whose entanglement
is described in Fig.~\ref{multienta3}, setting the evolution time to $t=5{\mu}{\rm s}$
and the heating rate to $\epsilon=200{\rm Hz}$ (resulting from a coupling to the bath
$\gamma=10^{-4} {\rm Hz}$ and a temperature $T=294^{\circ} {\rm K}$),
whose CM will be denoted by $\sig_3$.
The
family of (non-Gaussian) local, bounded, dichotomic observables
for the displaced parity test is given by
$\Pi_j(x_{j},p_{j})\equiv\hat{D}_j(x_j,p_j)^{\dag}(-1)^{\hat{n}_{j}}\hat{D}_{j}(x_j,p_j)$,
where $\hat{D}_{j}$ and $\hat{n}_j$ are
respectively the displacement and number
of phonons operators of ion $j$. The three observers, pertaining to
the three ions, randomly apply two different displacements
[$\hat{D}_j(x_j,p_j)$ and $\hat{D}_j(x'_j,p'_j)$] on their ions and
then measure parity locally.
Such a measurement is clearly non-Gaussian, and allows one to violate Bell inequalities with Gaussian states.
The expectation value of the operator
$\Pi(R)\equiv\Pi_1(x_{1},p_1)\otimes\Pi_2(x_{2},p_{2})\otimes\Pi_3(x_{3},p_{3})$
is simply proportional to the Wigner function $W(R)$ of the
composite system evaluated in the point $R=(x_1,x_2,x_3,p_1,p_2,p_3)^{\sf T}$:
$\langle\Pi(R)\rangle=(2/\pi)^3 W(R)$ \cite{wignerparity}.
For a three-mode Gaussian state with covariance matrix $\sig_3$ one has
$$W(R) = {\rm e}^{-\frac12R^{\sf T}\sig_{3}^{-1}R}/(\pi^3\sqrt{\det{\sig_{3}}}) \; .$$
The Bell-Klyshko inequality finally reads:
\bea
B_3&\equiv&\frac{8}{\pi^3}|W(x_1,x_2,x'_3,p_1,p_2,p'_3) + W(x_1,x'_2,x_3,p_1,p'_2,p_3) \nonumber \\
&&+ W(x'_1,x_2,x_3,p'_1,p_2,p_3) - W(x'_1,x'_2,x'_3,p'_1,p'_2,p'_3)|\le 2 \,.
\eea
Quantum mechanics allows for $2\le B_3\le4$.

\begin{figure}[t!]
\begin{tabular}{cc}
\includegraphics[width=6cm,height=6cm]{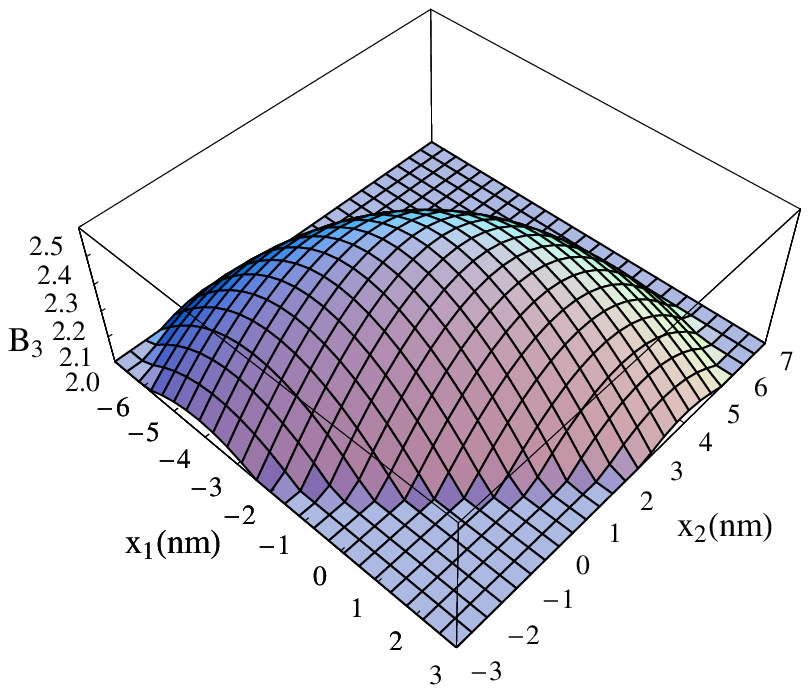}&
\includegraphics[width=6cm,height=6cm]{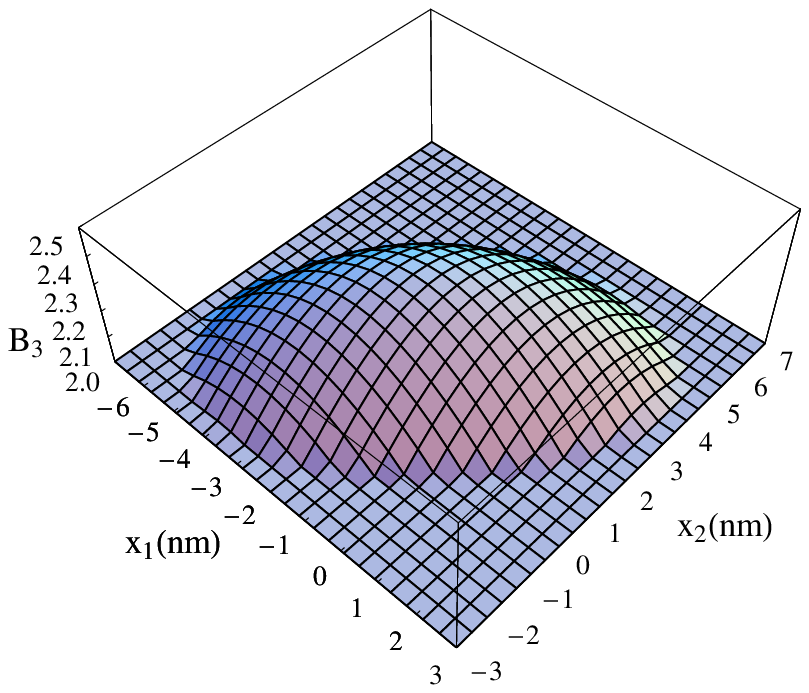}
\end{tabular}
\caption{Function $B_3$ for $p_1=p_2=p_3=p'_1=p'_2=p'_3=0$, $x'_1=x'_3= 6 {\rm nm}$,
$x'_2=-4 {\rm nm}$,
$-7 {\rm nm}\le x_1 = x_3 \le 3 {\rm nm}$ and $-3 {\rm nm}\le x_2 \le 7 {\rm nm}$
for the three-mode Gaussian state obtained as detailed in the caption of Fig.~\ref{multienta3}
after a time $t=5{\mu}{\rm s}$.
On the left hand-side no dissipation is accounted for, while on the right-hand side
a heating rate $\epsilon\simeq 200{\rm Hz}$ is introduced.
Dimensions were reintroduced assuming
$\omega_{L}=1{\rm MHz}$ and Ca ions.
The plane $B_3=2$, above which violation occurs, is the bottom of the plots;
in the displayed region, the functions reach maxima of $\simeq 2.45$
for the pure state case and of $\simeq 2.28$ with dissipation.}\label{viola3}
\end{figure}

Fig.~\ref{viola3} show a region in the space of displacements where the violation
of the inequality is substantial and remarkably stable.
It is apparent that, for such
choices of displacements, the tolerable error on the displacement operation needed to maintain the violation
is around $1.5 {\rm nm}$, which is within reach of current experimental capabilities.
Also, as shown in the plot on the right,
heating rates of $200{\rm Hz}$ are still compatible with a violation of the inequalities.

This preliminary study reveals very promising perspectives concerning
the violation of Bell inequalities with massive degrees of freedom.
Even if subject to a locality loophole, this endeavour
would still stand out as a major, not yet probed,
testing ground for fundamental quantum mechanics \cite{retzker05},
and epitomises the considerable promise radial modes hold for quantum
information and fundamental investigations alike.


\section{Summary and outlook}\label{conclu}

We demonstrated how the local
control of the trapping frequencies would allow
one to reproduce any linear optical manipulation on radial modes
of trapped ions.
Drawing from previous studies on similar settings, we pointed out
that phonon detection and homodyne
detection as well as the implementation of non-Gaussian operations
is possible in this setting. Next, we emphasized that, even
restricting to global control, such manipulations enjoy a high
efficiency in entanglement generation and a considerable resilience
in the face of currently achievable dissipation rates.
We then made manifest that radial modes could be used for the transmission
of quantum information, stored in finite dimensional subspaces of the bosonic
Hilbert space as well as in the full continuous variable domain.
Finally we showed that, through (achievable) measurements of
the displaced parity operator, Bell-like inequalities can be violated under
realistic decoherence and dissipation rates.
As a rule of thumb, applying to systems with $n\simeq6$ ions or less
and longitudinal trapping frequencies $\omega_{L}\simeq 1$ ${\rm MHz}$,
our study indicates that heating rates around $1$ ${\rm KHz}$ are sufficiently low for
coherent manipulations and robust entanglement generation in the continuous
variable regime, whereas non-locality tests and transmission of finite-dimensional
quantum dimensional information (stored in single phonons) are more delicate, requiring heating rates
around $100$ ${\rm Hz}$ to be carried out efficiently.

In the light of the above, the experimental pursuit
of the presented programme holds considerable promise, concerning
both technological developments, such as the storage and manipulation
of quantum information or the efficient generation of entanglement,
and tests of fundamental physical aspects, as in the nonlocality
test for massive degrees of freedom here discussed.
As a first step to push this analysis further,
a study is currently under way to examine the possibility to
swap the remarkable entanglement produced in the traps
to light modes, to ultimately beat parametric processes in the
generation of quantum optical CV entanglement \cite{noialtri}.

\ack
\noindent We thank F.G.S.L.~Brand\~ao, T.~Coudreau, H.~H\"affner,
M. Keller, W.~Lange, K.~Pregnell, D.M.~Segal and R.C.~Thompson
for helpful discussions. This work has been supported by the
European Commission under the Integrated Project QAP, funded by
the IST directorate as Contract Number015848, by an EU Marie
Curie IEF fellowship, by the EU STREP project CORNER, by the
Royal Society and is part of the EPSRC QIP-IRC and by EPSRC project number EP/E045049/1.

\end{document}